\documentclass[11pt,a4paper]{article}
\usepackage{amssymb}
\usepackage{amsmath}
\usepackage{amsfonts}
\usepackage{bbm}
\usepackage{amsthm}
\usepackage{mathrsfs}
\usepackage{hyperref}
\usepackage{marvosym}
\usepackage{color}
\usepackage[margin=2.6cm]{geometry}
\usepackage[all,cmtip]{xy}
\usepackage[utf8]{inputenc}

\definecolor{darkred}{rgb}{0.8,0.1,0.1}
\hypersetup{
     colorlinks=false,         
     linkcolor=darkred,
     citecolor=blue,
}

\theoremstyle{plain}

\theoremstyle{definition}

\numberwithin{equation}{section}

\def\nn{\nonumber}

\def\bbR{\mathbb{R}}
\def\bbC{\mathbb{C}}
\def\bbN{\mathbb{N}}
\def\bbZ{\mathbb{Z}}
\def\bbT{\mathbb{T}}

\def\ii{{\,{\rm i}\,}}

\def\Hom{\mathrm{Hom}}

\def\Imm{\mathrm{Im}}
\def\Ker{\mathrm{Ker}}

\def\id{\mathrm{id}}

\def\dd{\mathrm{d}}

\def\cc{\mathrm{c}}

\def\1{\mathbbm{1}}

\def\g{\mathfrak{g}}
\def\op{\mathrm{op}}

\newcommand{\ip}[2]{\big\langle #1,#2 \big\rangle}
\newcommand{\ips}[2]{\langle #1,#2\rangle}

\newcommand\mycom[2]{\genfrac{}{}{0pt}{}{#1}{#2}}

\def\DD{\mathsf{D}}

\def\FF{\mathfrak{F}}

\def\XX{\mathfrak{X}}
\def\YY{\mathfrak{Y}}

\def\Conf{\mathfrak{C}}
\def\Obs{\mathfrak{O}}
\def\Del{\mathfrak{Del}}

\def\Contr{\mathsf{Man}_{\text{\textcopyright}}}
\def\Man{\mathsf{Man}}
\def\Ch{\mathsf{Ch}}
\def\Ab{\mathsf{Ab}}

\def\colim{\mathrm{colim}}
\def\holim{\mathrm{holim}}
\def\hocolim{\mathrm{hocolim}}
\def\ext{\mathrm{ext}}

\def\sk{\vspace{2mm}}

\title{%
Homotopy colimits and global observables \\ in Abelian gauge theory
}

\author{%
Marco Benini$^{a}$, Alexander Schenkel$^b$ and Richard J.\ Szabo$^c$ \vspace{4mm}\\
{\small Department of Mathematics, Heriot-Watt University,}\\
{\small Colin Maclaurin Building, Riccarton, Edinburgh EH14 4AS, United Kingdom.}\vspace{1mm}\\
{\small Maxwell Institute for Mathematical Sciences, Edinburgh, United Kingdom.}\vspace{1mm}\\
{\small The Tait Institute, Edinburgh, United Kingdom.}\vspace{4mm}\\
 {\footnotesize \texttt{Email:} $^a$ \texttt{mbenini87@gmail.com} ~,~~ $^b$ \texttt{as880@hw.ac.uk} ~,~~$^c$  \texttt{R.J.Szabo@hw.ac.uk}}
 }

\date{May 2015}

\begin{document}

\maketitle

\begin{abstract}
We study chain complexes of field configurations and observables for Abelian gauge theory on contractible manifolds, and show that they can be extended to non-contractible manifolds by using techniques from homotopy theory. The extension prescription yields functors from a category of manifolds to suitable categories of chain complexes. The extended functors properly describe the global field and observable content of Abelian gauge theory, while the original gauge field configurations and observables on contractible manifolds are recovered up to a natural weak equivalence.
\end{abstract}

\paragraph*{Report no.:} EMPG--15--04
\paragraph*{Keywords:} Abelian gauge theory, global configurations and
observables, chain complexes, homotopy limits and colimits
\paragraph*{MSC 2010:} 70S15, 81T13, 55U15, 55N20



\bigskip

\tableofcontents

\bigskip

\section{Introduction and summary}

In a  classical field theory (without gauge symmetry) the field configurations
on a manifold $M$ are described in terms of the set of sections $\FF(M) := \Gamma^\infty(E_M)$
of some natural fibre bundle $E_M$ over $M$. For example, 
for a real scalar field theory the field configurations on $M$ 
are given by the set $C^\infty(M)$  of functions on $M$, which is the set of sections
of the trivial line bundle $\bbR\times M\to M$. Naturality allows us to regard 
$\FF : \Man^\op \to \mathsf{Sets}$ as a functor from the category of (say, oriented $m$-dimensional) 
manifolds to the category of sets. The {\em sheaf property} of the set of sections of a fibre bundle
allows us to capture all information about the global field configurations $\FF(M)$ on a manifold $M$
in terms of the local field configurations in an open cover $\{ U_i : i\in\mathcal{I}\}$ of $M$.
More precisely, the sheaf property says that the set $\FF(M)$ can be recovered (up to isomorphism) 
from any open cover $\{ U_i : i\in\mathcal{I}\}$ of $M$ by taking the limit 
\begin{flalign}\label{eqn:sheaf}
\xymatrix{
\FF(M)\ar[r]^-{\cong} & \lim\Big(\prod\limits_{i}\, \FF(U_i) \mathrel{\substack{\textstyle\rightarrow\\[-0.6ex]
                      \textstyle\rightarrow}} \prod\limits_{i,j}\, \FF(U_{ij})\Big)
}~
\end{flalign}
in the category $\mathsf{Sets}$. Here $\prod$ denotes the categorical 
product in $\mathsf{Sets}$ and as usual we denote the intersections by $U_{ij} := U_i\cap U_j$.
\sk

Another essential ingredient of a classical field theory is the characterization 
of the observables of the theory, which is usually done
by specifying for each manifold $M$ a suitable algebra  $\mathfrak{A}(M)$ of functions
on $\FF(M)$.\footnote{
The algebras $\mathfrak{A}(M)$ typically carry a Poisson structure for classical field theories 
defined on globally hyperbolic Lorentzian manifolds, see e.g.\ \cite{Brunetti:2012ar},
or they are noncommutative algebras after quantization, see e.g.\ \cite{Brunetti:2001dx}.
These additional structures will not play a role in the present paper because we do not
consider dynamical aspects of field theories or quantization.
}
Following \cite{Brunetti:2001dx}, an important guiding principle for the choice of the observable algebras $\mathfrak{A}(M)$
is the requirement of functoriality of the assignment $\mathfrak{A} : \Man \to \mathsf{Alg}$,
where $\mathsf{Alg}$ is a suitable category of algebras whose details depend on the context.
Another reasonable requirement 
is the {\em cosheaf property} of $\mathfrak{A}$, which would allow us to capture all information
about the global observables $\mathfrak{A}(M)$ on a manifold $M$
in terms of the local observables in an open cover $\{ U_i : i\in\mathcal{I}\}$ of $M$.
More precisely, the cosheaf property demands that the algebra $\mathfrak{A}(M)$ can be recovered
(up to isomorphism) from any open cover $\{ U_i : i\in\mathcal{I}\}$ of $M$
by taking the colimit
\begin{flalign}\label{eqn:cosheaf}
\xymatrix{
\mathfrak{A}(M) &\ar[l]_-{\cong}  \colim\Big( \coprod\limits_{i,j}\, \mathfrak{A}(U_{ij})
\mathrel{\substack{\textstyle\rightarrow\\[-0.6ex]
                      \textstyle\rightarrow}}
\coprod\limits_{i}\, \mathfrak{A}(U_i) \Big)
}~
\end{flalign}
in the category $\mathsf{Alg}$. Here $\coprod$ denotes the categorical coproduct in $\mathsf{Alg}$.
\sk

When studying explicit examples of classical (and also quantum) field theories
it might very well happen that one can construct rather easily the field configurations
$\FF(M)$ and the observables $\mathfrak{A}(M)$  of the theory on a special
class of manifolds,  e.g.\ on the category of contractible manifolds $\Contr$,
but that the construction becomes much more involved for non-contractible
manifolds. Reasons for this might be global features, such as non-trivial bundles and
topological charges, which are related to topologically non-trivial manifolds.
In such a situation one would obtain two functors $\FF : \Contr^\op \to \mathsf{Sets}$
and $\mathfrak{A} : \Contr \to \mathsf{Alg}$ describing the field configurations
and observables of the theory only on contractible manifolds, and the goal is to then
`extend' these functors to the category of all manifolds $\Man$.
In view of the desired sheaf and cosheaf properties, a
reasonable procedure for obtaining such extensions is to {\em define} the field configurations
$\FF(M)$ and the observables $\mathfrak{A}(M)$ on a generic manifold $M$
in terms of the limit or, respectively, the colimit of a diagram induced by
a suitable open contractible cover of $M$. In the context of algebraic quantum field theory,
such a procedure has been proposed by Fredenhagen and it is called
the `universal algebra', see e.g.\ \cite{Fre90,Fre93,FRS92}. 
\sk

In gauge theories the structures discussed above become considerably more complicated. First of all,
the gauge field configurations on a manifold $M$ are {\em not} described by a set,
but by a groupoid. For example, the field configurations of gauge theory with structure group $G$
on a manifold $M$ are described by the groupoid with objects given by all principal $G$-bundles over $M$ endowed with a connection 
and morphisms given by all principal $G$-bundle isomorphisms compatible with the connections (i.e.\ gauge transformations).
Instead of forming a sheaf, the collection of these groupoids for all manifolds $M$ forms
a stack, see e.g.\ \cite{Fantechi,Vistoli} for an introduction.
For our purposes, a more explicit and also more suitable characterization of stacks in terms of homotopy sheaves of
groupoids has been developed by Hollander \cite{Hollander}. A stack is the same thing
as a functor $\mathfrak{G} : \Man^\op \to\mathsf{Groupoids}$ to the category of groupoids
which satisfies the {\em homotopy sheaf property}, i.e.\ for any manifold $M$ 
the groupoid $\mathfrak{G}(M)$ can be recovered (up to weak equivalence) 
from any open cover $\{U_i : i\in\mathcal{I}\}$ of $M$ by taking the {\em homotopy limit}
\begin{flalign}
\xymatrix{
\mathfrak{G}(M)\ar[r]^-{\sim} & \holim\Big(\prod\limits_{i}\, \mathfrak{G}(U_i) \mathrel{\substack{\textstyle\rightarrow\\[-0.6ex]
                      \textstyle\rightarrow}} \prod\limits_{i,j}\, \mathfrak{G}(U_{ij}) \mathrel{\substack{\textstyle\rightarrow\\[-0.6ex]
                      \textstyle\rightarrow \\[-0.6ex]
                      \textstyle\rightarrow}} \prod\limits_{i,j,k}\, \mathfrak{G}(U_{ijk}) 
\mathrel{\substack{\textstyle\rightarrow\\[-0.6ex]
                      \textstyle\rightarrow \\[-0.6ex]
			   \textstyle\rightarrow \\[-0.6ex]
                      \textstyle\rightarrow}}
\cdots \Big)
}
\end{flalign}
in the category $\mathsf{Groupoids}$.\footnote{For a concise and very readable introduction to homotopy theory and
model categories see \cite{Dwyer}.} Note that forming the coarse moduli spaces 
(i.e.\ the gauge orbit spaces) of a homotopy sheaf in general does not yield a sheaf. Hence the groupoid
point of view is essential for `gluing' local gauge field configurations to global ones.
\sk

Classical observables for gauge theories may be described
by taking suitable `function algebras' on groupoids, which can be modeled
by cosimplicial algebras or differential graded algebras, see Section \ref{sec:contractible} below for details.
A natural requirement for the choice of such `function algebras' is again
functoriality, i.e.\ we seek a functor $\mathfrak{B} : \Man \to \mathsf{cAlg}$
to the category of cosimplicial algebras (or a functor $\mathfrak{B} : \Man \to \mathsf{dgAlg}$
to the category of differential graded algebras). 
Instead of the cosheaf property which appears in ordinary field theories,
this functor should satisfy the {\em homotopy cosheaf property}, i.e.\ the cosimplicial (or differential graded)
algebra $\mathfrak{B}(M)$ can be recovered
(up to weak equivalence) from any open cover $\{ U_i : i\in\mathcal{I}\}$ of $M$
by taking the {\em homotopy colimit}
\begin{flalign}\label{eqn:homotopycosheaf}
\xymatrix{
\mathfrak{B}(M) &\ar[l]_-{\sim}  \hocolim\Big(\cdots
\mathrel{\substack{\textstyle\rightarrow\\[-0.6ex]
                      \textstyle\rightarrow \\[-0.6ex]
			   \textstyle\rightarrow \\[-0.6ex]
                      \textstyle\rightarrow}}
\coprod\limits_{i,j,k} \, \mathfrak{B}(U_{ijk})\mathrel{\substack{\textstyle\rightarrow\\[-0.6ex]
                      \textstyle\rightarrow \\[-0.6ex]
                      \textstyle\rightarrow}}\coprod\limits_{i,j}\, \mathfrak{B}(U_{ij})
\mathrel{\substack{\textstyle\rightarrow\\[-0.6ex]
                      \textstyle\rightarrow}}
\coprod\limits_{i}\, \mathfrak{B}(U_i) \Big)
}
\end{flalign}
in the category $\mathsf{cAlg}$ (or in the category $\mathsf{dgAlg}$).\footnote{See \cite{CC,Jardine} for details on the relevant model category structures on 
$\mathsf{cAlg}$ and $\mathsf{dgAlg}$.}
\sk

In gauge theories we are exactly in the situation where the groupoids of field configurations
are rather simple and explicit for contractible manifolds $M$, while they are much 
harder to describe for non-contractible manifolds. This is because on a contractible manifold $M$
all principal $G$-bundles are isomorphic to the trivial principal $G$-bundle $M\times G\to M$.
Consequently, gauge field observables on contractible manifolds are also much easier to describe
than those on non-contractible manifolds. Hence we can rather easily get
two functors $\mathfrak{G} : \Contr^\op \to \mathsf{Groupoids}$
and $\mathfrak{B} : \Contr \to \mathsf{cAlg}$ (or $\mathfrak{B} : \Contr \to \mathsf{dgAlg}$) 
which describe gauge field configurations and observables only on contractible manifolds. 
The goal is then to extend these two functors to the category $\Man$
by taking homotopy limits or, respectively, homotopy colimits of diagrams induced
by suitable open contractible covers of manifolds $M$. 
From the perspective of algebraic quantum field theory,
these constructions may be interpreted as a gauge theoretic (or homotopy theoretic) 
version of Fredenhagen's `universal algebra' construction.
Let us emphasize again the importance of describing gauge field
configurations in terms of groupoids and observables in terms of cosimplicial (or differential graded) algebras,
instead of working with gauge orbit spaces and gauge invariant observable algebras.
As an explicit example of what may go wrong when not doing so, 
see \cite{Dappiaggi:2011zs,Fewster:2014hba} where the `universal algebra' has been constructed for
{\em gauge invariant} observable algebras of Abelian gauge theory. These `universal algebras' fail to 
produce the correct global gauge invariant observable algebras \cite{Becker:2014tla} because they neglect flat 
connections and violate the quantization condition for magnetic
charges in the integral cohomology $H^2(M,\bbZ)$.
See also \cite{Benini:2013ita, Benini:2013tra} for a presentation 
of the global gauge invariant observable algebras for a fixed but arbitrary principal bundle
and \cite{Fewster:2003ey,Dappiaggi:2011cj,Sanders:2012sf,Ciolli:2013pta,Benini:2014vsa} for 
the trivial principal bundle. Certain aspects of non-Abelian gauge theories and also gravity 
in this context have been studied in~\cite{Hollands:2007zg,Fredenhagen:2011mq,Ciolli:2011xv,Brunetti:2013maa,Khavkine:2014kya,Khavkine:2015fwa}.
\sk

In this paper we shall make explicit and test the above ideas for constructing global gauge field configurations
and observables by homotopy theoretic techniques. We shall consider the simplest example 
of a classical gauge theory, namely that whose structure group is the
circle group $G = \bbT = U(1)$. From the
perspective of differential cohomology, there already exist several
models for the groupoid of gauge potentials in this case which have
been discussed from the perspective of `locality' of (generalized)
Abelian gauge theories: The category of
differential cocycles constructed by~\cite{HS} is based on singular
cochains (see also~\cite[Section~2.4]{Szabo:2012hc}), while the
\v{C}ech theoretic model of~\cite{Freed:1999vc} is somewhat closer in
spirit to our approach (see also~\cite[Example 1.11]{Freed:2000ta}); see also~\cite{Belov:2006jd} for a more
heuristic proposal. A similar functorial description of abelian gauge theories is discussed by~\cite{Fiorenza:2013jz}.
To simplify our constructions, we shall study this gauge theory on a purely kinematical level, 
leaving out both dynamical aspects (i.e.\ Maxwell's equations) and
quantization for the moment.
\sk

The outline of the remainder of this paper is as follows: In Section~\ref{sec:contractible}
we give an explicit and very useful description of the groupoids of gauge field configurations
on contractible manifolds in terms of chain complexes of Abelian groups.
This formulation allows us to identify a simple class of gauge field observables,
given by smooth group characters,
which also forms a chain complex of Abelian groups.
Our constructions are functorial in the sense that we obtain a functor
$\Conf : \Contr^\op \to \Ch_{\geq 0}(\Ab)$ describing gauge field configurations
and a functor $\Obs : \Contr \to \Ch_{\leq 0}(\Ab)$ describing observables
on the category of contractible manifolds $\Contr$.
In Section~\ref{sec:globalconf} we shall construct an extension
of the functor $\Conf : \Contr^\op \to \Ch_{\geq 0}(\Ab)$ to the
category $\Man$ of all manifolds by using homotopy limits.
For this we first show that any manifold $M$ has a canonical open cover by contractible subsets,
which induces a canonical diagram of gauge field configurations on $M$.
We compute explicitly the homotopy limit of this diagram and show that
it is isomorphic to the Deligne complex in the canonical cover.
This will imply that our homotopy limit describes all possible gauge field configurations
on $M$, including also non-trivial principal $\bbT$-bundles whenever $H^2(M,\bbZ)\neq 0$.
As the canonical cover is functorial, it is easy to prove that the global field configurations
given by the homotopy limits are described by a functor $\Conf^\ext : \Man^\op \to \Ch_{\geq 0}(\Ab)$.
We shall show that this functor is an extension (up to a natural quasi-isomorphism) 
of our original functor $\Conf : \Contr^\op \to \Ch_{\geq 0}(\Ab)$.
In Section~\ref{sec:globalobservables} we shall focus on the gauge field observables
and construct an extension of the functor $\Obs : \Contr \to \Ch_{\leq 0}(\Ab)$ to the category
$\Man$  by using homotopy colimits. Similarly to the gauge field configurations, we obtain a canonical diagram
of gauge field observables on any manifold $M$ and we compute explicitly the homotopy colimits. 
Functoriality of the global observables $\Obs^\ext : \Man \to \Ch_{\leq 0}(\Ab)$ is again a simple consequence
of functoriality of the canonical cover. We then show that this functor is an extension (up to a natural quasi-isomorphism) 
of our original functor $\Obs : \Contr \to \Ch_{\leq 0}(\Ab)$. Finally, we construct
a natural pairing between global gauge field configurations and observables, 
which allows us to show that our class of observables separates all possible gauge field configurations. 
Two appendices at the end of the paper summarize some of the more technical
details which are used in the main text. In Appendix~\ref{app:DoldKan} we review
the (dual) Dold-Kan correspondence, which is an important technical tool for our constructions.
In Appendix~\ref{app:holimcolim} we summarize the explicit techniques to compute homotopy (co)limits
for chain complexes of Abelian groups given in~\cite[Section 16.8]{Dugger} and~\cite{RodGon}.


\section{\label{sec:contractible}Local gauge field configurations and observables}

In this section we consider gauge fields on contractible manifolds; in
this paper all manifolds considered are oriented.

\subsection{Groupoids and cosimplicial algebras}

Let $G$ be a (matrix) Lie group, 
$\g$ its Lie algebra and $M$ a {\em contractible}  manifold.
Then all principal $G$-bundles over $M$ are 
isomorphic to the trivial $G$-bundle, and the field configurations of gauge theory with structure group $G$ on $M$
are described by the $\g$-valued one-forms $\Omega^1(M,\g)$; elements $A\in\Omega^1(M,\g)$
are typically called `gauge potentials'. Recall that gauge theory
comes with a distinguished notion of gauge group, the group of vertical automorphisms
of the principal $G$-bundle. In the present case the gauge group
may be identified with the group of $G$-valued smooth functions $C^\infty(M,G)$
and it acts on gauge potentials from the left via 
\begin{flalign}\label{eqn:gaugeaction}
\rho : C^\infty(M,G)\times \Omega^1(M,\g)\longrightarrow \Omega^1(M,\g)~, \qquad (g,A)\longmapsto 
\rho(g,A) = g\,A\,g^{-1} + g\,\dd g^{-1}~,
\end{flalign}
where $\dd$ denotes the exterior derivative.
\sk

Having available both gauge potentials and gauge transformations, one
can ask which mathematical structure is suitable for describing the
relevant field content of gauge 
theory on $M$. The most obvious option is to take the orbit space $\Omega^1(M,\g)/C^\infty(M,G)$ 
under the $\rho$-action, which identifies
all gauge potentials that differ by a gauge transformation; this is often called the `gauge orbit space'.
However, there are several problems with the orbit space construction: First, 
even though both $\Omega^1(M,\g)$ and $C^\infty(M,G)$ can be equipped with
a suitable (locally convex infinite-dimensional) smooth manifold structure, the orbit space is in general 
singular~\cite{Abbati:1986ww,ACM}. Second, forming the orbit space inevitably leads to a substantial loss of information;
even though we can still decide whether or not two gauge potentials $A$ and $A^\prime$ are gauge equivalent,
we cannot keep track of the gauge transformation $g$ that identifies $A$ with $A^\prime$ when they are equivalent.
The latter information is essential whenever one wants to obtain global field configurations of gauge theory on a {\em topologically
non-trivial} manifold $M$ by `gluing' local configurations in contractible patches. 
A classic example is Dirac's famous magnetic monopole which represents the Chern class in Abelian gauge 
theory with structure group the circle group $G=\bbT=U(1)$: Its construction is based on gauge potentials $A_1$ and $A_2$ on an open cover $\{ U_1,U_2\}$ of a topologically non-trivial manifold $M$ 
subject to the requirement $A_2\vert_{U_{12}} - A_{1}\vert_{U_{12}} = g_{12}\, \dd g_{12}^{-1}$
for some fixed $g_{12}\in C^\infty(U_{12},\bbT)$ on the overlap $U_{12} = U_{1}\cap U_2$.
This operation of `gluing up to gauge transformations' cannot be described in terms of gauge orbit spaces.
\sk

In order to solve these and other problems, a more modern perspective suggests that,
instead of looking at gauge orbits, one should organize the gauge potentials and gauge transformations
into a groupoid. Recall that a groupoid is a small category in which every morphism is invertible.
The groupoid corresponding to gauge theory with structure group $G$ on a contractible
manifold $M$ is simply the action groupoid $C^\infty(M,G)\ltimes
\Omega^1(M,\g) \rightrightarrows \Omega^1(M,\g)$: The set of objects is $\Omega^1(M,\g)$ and the set of morphisms is $C^\infty(M,G)\times \Omega^1(M,\g)$,
an element $(g,A)$ of which should be interpreted as an arrow 
\begin{flalign}
\rho(g,A) \xleftarrow{ \ (g,A) \ } A
\end{flalign}
starting at $A$ and ending at the gauge transform (\ref{eqn:gaugeaction}) of $A$ by $g$.
Two morphisms $(g^\prime,A^\prime)$ and $(g,A)$ are composable whenever $A^\prime = \rho(g,A)$
and the composition reads as $(g^\prime,A^\prime)\circ(g,A) = (g^\prime\,g,A)$. The identity morphisms
are $\id_A = (e,A)$, where $e$ is the identity element in $C^\infty(M,G)$, i.e.\ the constant function
to the identity element of the structure group $G$. As an aside, note that one can use the techniques 
of \cite{Abbati:1986ww} to realize that this action groupoid is moreover a (locally convex infinite-dimensional) Lie groupoid,
i.e.\ a groupoid carrying a smooth structure. This smooth structure will not play a role in the present paper, since
we will shortly restrict ourselves to Abelian gauge theory with structure group $G=\bbT$, 
which can be studied in purely algebraic terms.
However, in studies of non-Abelian gauge theory the smooth structures
will play an important role.
\sk

Instead of using groupoids, we may equivalently organize the gauge potentials and gauge transformations into a simplicial set
(or even a simplicial manifold if we use the smooth structure discussed above).
Recall that a simplicial set is a collection $\{S_n\}_{n\in\bbN_0}$ of sets together with
face maps $\partial_i^n : S_n\to S_{n-1}$, for $n\geq 1$ and $i=0,1,\dots,n$, and degeneracy maps
$\epsilon_i^n : S_{n} \to S_{n+1}$, for $n\geq 0$ and $i=0,1,\dots,n$, satisfying simplicial identities, see 
e.g.~\cite[Section I.1]{GJ99}.
The simplicial set corresponding to our groupoid (which is called its nerve) may be depicted as
\begin{flalign}\label{eqn:nerveYM}
\xymatrix{
\Omega^1(M,\g) & \ar@<0.5ex>[l]\ar@<-0.5ex>[l] C^\infty(M,G)\times \Omega^1(M,\g) &\ar[l]\ar@<1ex>[l]\ar@<-1ex>[l] 
C^\infty(M,G)^{\times 2}\times \Omega^1(M,\g) & \ar@<0.5ex>[l]\ar@<-0.5ex>[l]\ar@<1.5ex>[l]\ar@<-1.5ex>[l]  \cdots
}~,
\end{flalign}
where the arrows are the face maps and we have suppressed the degeneracy maps for notational convenience. 
Explicitly, the face maps read as
\begin{subequations}\label{eqn:facedegYM}
\begin{flalign}
\nn \partial_i^n : C^\infty(M,G)^{\times n} \times \Omega^1(M,\g) &\longrightarrow C^\infty(M,G)^{\times n-1}\times \Omega^1(M,\g) ~,~~\\
(g_1,\dots, g_n,A ) &\longmapsto \begin{cases}
(g_2,\dots,g_n, A) & \text{~,~~for }i=0~,\\
(g_1,\dots,g_i\,g_{i+1},\dots g_n,A) &\text{~,~~for }i=1,\dots,n-1~,\\
(g_1,\dots,g_{n-1}, \rho(g_n,A)) &\text{~,~~for }i=n~,
\end{cases}
\end{flalign}
and the degeneracy maps read as
\begin{flalign}
\nn \epsilon_i^n : C^\infty(M,G)^{\times n} \times \Omega^1(M,\g) &\longrightarrow C^\infty(M,G)^{\times n+1}\times \Omega^1(M,\g) ~,~~\\
(g_1,\dots, g_n,A ) &\longmapsto (g_1,\dots,g_i,e,g_{i+1},\dots,g_{n},A)~.
\end{flalign}
\end{subequations}

The simplicial set perspective has the advantage of making clear how to
describe gauge theory observables. Interpreting
(\ref{eqn:nerveYM}) as the simplicial set of gauge field configurations,
it is natural to model (classical) observables as functions on it. This can be done by 
taking the algebra of complex-valued functions $C(\,-\,,\bbC)$ on each degree of (\ref{eqn:nerveYM}). 
Doing so, a cosimplicial algebra is obtained by dualizing the face and degeneracy maps 
to co-face and co-degeneracy maps under the {\em contravariant} functor
$C(\,-\,,\bbC) : \mathsf{Sets}\to \mathsf{Alg}$ between sets and algebras.
Restricting to infinitesimal gauge transformations, this picture
reduces nicely to the well-known BRST formalism, 
see \cite{Fredenhagen:2011an} for a presentation of this topic 
in the context of the algebraic approach to field theory.
By the dual Dold-Kan correspondence (see Appendix~\ref{app:DoldKan}) we can regard our cosimplicial algebra as 
a differential graded algebra (dg-algebra), see also \cite{CC} for more details. 
This dg-algebra can be `linearized' via a procedure called the van-Est map 
(here we require the smooth structure mentioned above) to yield the
BRST algebra (Chevalley-Eilenberg dg-algebra) 
of gauge theory, see e.g.\ \cite{Crainic}. 
It is important to stress that this linearization procedure neglects finite gauge transformations
and hence leads to an incomplete description of gauge theory.
Our cosimplicial algebra (or its associated dg-algebra) should be interpreted as an improvement
of the usual BRST algebra, which keeps track of all gauge transformations and not only of the infinitesimal ones; 
in fact, finite gauge transformations are essential for gluing local field configurations to global ones. 
Using the analogy with the BRST formalism, we may call the factors $C^\infty(M,G)$ in (\ref{eqn:nerveYM})
the `ghost fields'. Notice that these ghost fields are non-linear in the sense that they are functions with values in the
structure group, while the ordinary ghost fields in the BRST formalism are described by the tangent space $C^\infty(M,\g)$
at the identity $e\in C^\infty(M,G)$ and hence they are linear.

\subsection{Abelian gauge theory}

In the remainder of this paper we shall fix the structure group
$G=\bbT$ with Lie algebra $\mathfrak{t}=\ii \bbR$ and hence consider only Abelian gauge theory. Then the
structures described above simplify considerably.
In particular, all sets appearing in (\ref{eqn:nerveYM}) naturally become Abelian groups with respect to the 
direct product group structure given by
\begin{flalign}
(g_1,\dots,g_n,A)\,(g_1^\prime,\dots,g_n^\prime,A^\prime) := 
(g_1\,g_1^\prime,\dots, g_n\,g_n^\prime,A+A^\prime)~.
\end{flalign}
Moreover, the action of the gauge group on gauge potentials (\ref{eqn:gaugeaction})
simplifies to $\rho(g,A) = A + g\,\dd g^{-1}$, and it is easy to show that the face and degeneracy maps (\ref{eqn:facedegYM})
are group homomorphisms. It follows that the simplicial set (\ref{eqn:nerveYM}) is a simplicial Abelian group,
which under the Dold-Kan correspondence can be identified with a non-negatively graded chain complex of Abelian groups,
see Appendix \ref{app:DoldKan}.
This chain complex is called the normalized Moore complex
and in the present case it reads explicitly as
\begin{flalign}\label{eqn:chainYM}
\Conf(M) := \Big(\bigoplus_{n\geq 0}\, \Conf(M)_n \, ,~\delta \Big) := \Big(\Omega^1(M,\mathfrak{t})\oplus C^\infty(M,\bbT) \, ,~\delta \Big)~,
\end{flalign}
where $\Omega^1(M,\mathfrak{t})$ sits in degree $0$ and $C^\infty(M,\bbT)$ sits in degree $1$.
As a convenient sign convention, we shall take as the
differential (of degree $-1$) in $\Conf(M)$ the negative of the differential in the normalized Moore complex 
(\ref{eqn:Moorecomplex}), i.e.\
\begin{flalign}\label{eqn:chainYMdifferential}
\delta (A\oplus g) = (g\,\dd g^{-1}) \oplus 0~.
\end{flalign}
$M$ being contractible, the homology $H_\ast$ of the chain complex $\Conf(M)$ is given by
\begin{flalign}\label{eqn:chainYMhom}
 H_{0}(\Conf(M)) = \frac{\Omega^1(M,\mathfrak{t})}{\dd C^\infty(M,\mathfrak{t})}~,\qquad H_{1}(\Conf(M)) \simeq \bbT ~,
\end{flalign}
which gives the gauge orbit space in degree $0$ and the global constant gauge transformations in degree $1$. 
Notice that the first homology group is not a vector space, but only 
an Abelian group. This feature naturally distinguishes between the Abelian gauge theories 
with structure groups $G = \bbT$ and $G = \bbR$: Both theories have the same zeroth homology (i.e.\ the same 
gauge orbit space) on contractible manifolds, but they differ in the first homology which is always isomorphic to $G$.
\sk

For Abelian gauge theory we also obtain a distinguished class of observables:
Since in this case (\ref{eqn:nerveYM}) is a simplicial Abelian group, 
instead of all complex-valued functions $C(\,-\,,\bbC)$ on each degree, 
we can take only those functions which are group
characters, i.e.\ homomorphisms of Abelian groups $\Hom_{\mathsf{Ab}}(\,-\,,\bbT)$ to the circle group.
The group characters do not form an algebra, but rather an Abelian
group called the character group; of course
one can generate an algebra by the group characters, but this will not be done in the present paper.
The character group should be interpreted as a generalization of the vector space of linear observables
for a real scalar field theory, which also does not form an algebra, but which generates a polynomial algebra.
Taking the character groups $\Hom_{\mathsf{Ab}}(\,-\,,\bbT)$ in each degree of (\ref{eqn:nerveYM})
gives rise to a cosimplicial Abelian group because all face and degeneracy maps dualize to co-face and
co-degeneracy maps. Under the dual Dold-Kan correspondence this can be identified with a non-positively graded
chain complex of Abelian groups, see Appendix \ref{app:DoldKan}, which for our model explicitly reads as
\begin{flalign}\label{eqn:obschainYMpre}
\Big(C^\infty(M,\bbT)^\ast\oplus\Omega^1(M,\mathfrak{t})^\ast \, ,~\delta^\ast\Big)~, 
\end{flalign}
where ${\,-\,}^{\ast}:= \Hom_{\mathsf{Ab}}(\,-\,,\bbT)$. Here $C^\infty(M,\bbT)^\ast$ sits in degree $-1$ and
$\Omega^1(M,\mathfrak{t})^\ast$ sits in degree $0$, while the differential $\delta^\ast$  (of degree $-1$) 
is the dual of the differential (\ref{eqn:chainYMdifferential}). Using the smooth character groups as in~\cite{Becker:2014tla},
the chain complex (\ref{eqn:obschainYMpre}) can be restricted to
\begin{flalign}\label{eqn:obschainYM}
\Obs(M) := \Big(\bigoplus_{n\leq 0}\, \Obs(M)_n \, ,~\delta^\ast\Big) 
:=\Big( \Omega^m_{\cc,\bbZ}(M)\oplus \Omega^{m-1}_\cc(M) \, ,~\delta^\ast\Big)~,
\end{flalign}
where $m$ is the dimension of $M$ and the subscript ${}_{\cc}$ indicates differential forms 
of compact support. By $\Omega^m_{\cc,\bbZ}(M)$
we denote the subgroup of $\Omega^m_{\cc}(M)$ consisting of compactly supported top-degree forms 
which integrate to an integer, i.e.\ $\chi\in \Omega^m_{\cc,\bbZ}(M)$
if and only if $\int_M \, \chi \in\bbZ$.
It is instructive to explain in more detail how (\ref{eqn:obschainYM}) defines group characters on 
(\ref{eqn:chainYM}) and to give an explicit formula for $\delta^\ast$: Let us define the non-degenerate pairing
\begin{flalign}
\nn \ip{-}{-}_M : \Obs(M)\times \Conf(M) &\longrightarrow \bbT~,~~\\
(\chi\oplus \varphi,A\oplus g) &\longmapsto
\exp\Big(\int_M\, \big(A\wedge \varphi + \log(g)\,\chi\big)\Big)~.\label{eqn:pairing}
\end{flalign}
We observe that (\ref{eqn:pairing}) is a bi-character, 
i.e.\ 
\begin{subequations}
\begin{flalign}
\ip{\chi \oplus \varphi + \chi^\prime\oplus \varphi^\prime}{A\oplus g}_M &= \ip{\chi \oplus \varphi }{A\oplus g}_M~\ip{\chi^\prime \oplus \varphi^\prime }{A\oplus g}_M~,\\[4pt]
\ip{\chi \oplus \varphi}{A\oplus g + A^\prime \oplus g^\prime}_M &=
\ip{\chi \oplus \varphi}{A\oplus g}_M ~ \ip{\chi \oplus \varphi}{A^\prime \oplus g^\prime}_M~,
\end{flalign}
\end{subequations}
and that it is compatible with the gradings of $\Obs(M)$  and $ \Conf(M)$ if we
take the target $\bbT$ to sit in degree $0$.
The differential $\delta^\ast$ in $\Obs(M)$ is defined via the duality induced by  (\ref{eqn:pairing});
we compute
\begin{flalign}
\nn \ip{\chi \oplus \varphi}{\delta(A\oplus g)}_M &= \ip{\chi \oplus \varphi}{(g\,\dd g^{-1})\oplus 0)}_M \\[4pt]
\nn &=\exp\Big(-\int_M \, \dd \log(g)\wedge \varphi \Big) \\[4pt]
&= \ip{\dd\varphi\oplus 0}{A\oplus g}_M =: \ip{\delta^\ast(\chi\oplus \varphi)}{A\oplus g}_M~,
\end{flalign}
from which we find $\delta^\ast(\chi \oplus\varphi) = \dd\varphi\oplus 0$. 
Recalling that in the present section all manifolds are contractible, 
it follows that the homology $H_\ast$ of the chain complex (\ref{eqn:obschainYM})
is given by
\begin{flalign}
H_{-1}(\Obs(M)) \simeq \bbZ~,\qquad H_0(\Obs(M)) = \Omega^{m-1}_{\cc,\dd}(M):=\Ker\big(\dd :\Omega^{m-1}_{\cc}(M) \to\Omega^m_\cc(M) \big)~.
\end{flalign}
Comparing these groups with (\ref{eqn:chainYMhom}), we see that $H_{-1}(\Obs(M))$ contains exactly the group 
characters on $H_{1}(\Conf(M))$ and that $H_0(\Obs(M))$ is given by the (smooth) group characters on the gauge orbit
space $H_0(\Conf(M))$, i.e.\ the zeroth homology of $\Obs(M)$ describes gauge invariant group characters and hence
gauge invariant observables of that kind.
\sk

All of these constructions are functorial. Let us denote by $\Contr$ the following category of contractible manifolds:
The objects in $\Contr$ are all contractible oriented manifolds of a fixed dimension $m$ (which we suppress from the notation)
and the morphisms in $\Contr$ are all orientation preserving open embeddings.
The chain complexes in (\ref{eqn:chainYM}) are described by a functor
$\Conf : \Contr^\op \to \Ch_{\geq 0}(\Ab)$, where $\Contr^\op$ denotes the opposite category (i.e.\ the category with
reversed arrows) of $\Contr$ and $\Ch_{\geq 0}(\Ab)$ is the category of non-negatively graded chain 
complexes of Abelian groups; the functor $\Conf$ assigns to any object $M$ in $\Contr$
the chain complex $\Conf(M)$ given in (\ref{eqn:chainYM}) and to any morphism $f^\op : M^\prime\to M$ in $\Contr^\op$ (i.e.\
any morphism $f : M\to M^\prime$ in $\Contr$)  the morphism of chain complexes
\begin{flalign}\label{eqn:Confpullback}
\Conf(f^{\op}) := f^\ast : \Conf(M^\prime)\longrightarrow \Conf(M) ~,
\qquad A^\prime\oplus g^\prime\longmapsto f^\ast(A^\prime)\oplus f^\ast(g^\prime)~,
\end{flalign}
where $f^\ast$ is the pull-back of functions/differential forms along $f$.
Similarly, the chain complexes in (\ref{eqn:obschainYM}) are described by a functor $\Obs : \Contr \to \Ch_{\leq 0}(\Ab)$
to the category of non-positively graded chain complexes of Abelian
groups;
the functor $\Obs$ assigns to any object $M$ in $\Contr$ the chain complex $\Obs(M)$ given in (\ref{eqn:obschainYM})
and to any morphism $f : M\to M^\prime$ in $\Contr$ the morphism of chain complexes
\begin{flalign}\label{eqn:Obspushforward}
\Obs(f) :=f_\ast : \Obs(M)\longrightarrow \Obs(M^\prime)~, \qquad \chi\oplus \varphi\longmapsto f_\ast(\chi)\oplus f_\ast(\varphi)~,
\end{flalign}
where $f_\ast$ is the push-forward of compactly supported differential forms along $f$.
It follows that the pairing (\ref{eqn:pairing}) is natural in the sense that the diagram
\begin{flalign}
\xymatrix{
\ar[d]_-{f_\ast\times \id}\Obs(M) \times \Conf(M^\prime) \ar[rr]^-{\id \times f^\ast} && \Obs(M)\times \Conf(M)\ar[d]^-{\ips{-}{-}_M}\\
\Obs(M^\prime)\times \Conf(M^\prime)\ar[rr]_-{\ips{-}{-}_{M^\prime}}&&\bbT
}
\end{flalign}
commutes for all morphisms $f: M\to M^\prime$ in $\Contr$.


\section{\label{sec:globalconf}Homotopy limits and global gauge field configurations}
Our functor $\Conf : \Contr^\op \to \Ch_{\geq 0} (\Ab)$ given by (\ref{eqn:chainYM})
and (\ref{eqn:Confpullback}) describes the chain complexes of gauge field configurations
(together with gauge transformations) of Abelian gauge theory on {\em contractible}  manifolds.
Notice that for an arbitrary manifold $M$ the chain complex in (\ref{eqn:chainYM}) does not
necessarily describe all gauge field configurations on $M$: 
For example, $M$ might have a non-trivial second cohomology $H^2(M,\bbZ)\neq 0$, 
hence allowing for non-trivial principal $\bbT$-bundles which are not captured by (\ref{eqn:chainYM}). 
The goal of this section is to extend the functor  $\Conf : \Contr^\op \to \Ch_{\geq 0} (\Ab)$ 
to the category $\Man$ of {\em all} oriented $m$-dimensional manifolds (with
morphisms given by orientation preserving open embeddings) by using 
homotopy theoretic techniques.

\subsection{\label{subsec:canonicalcover}Canonical diagrams of gauge field configurations}
To any object $M$ in $\Man$ we can assign the category $\DD(M)$ of all {\em contractible} open subsets
$U\subseteq M$ with morphisms given by subset inclusions $U\subseteq V$. 
The set of objects $\DD(M)_0$ of $\DD(M)$ is an open cover of $M$, i.e.\ $\bigcup_{U\in \DD(M)_0}\, U = M$.
Notice that, even though every $U\in \DD(M)_0$ is contractible, the
open cover $\DD(M)_0$ is not a good cover: The intersection of two contractible open subsets fails to be contractible in general. 
This is not problematic since our main constructions do not use intersections.
To any morphism $f  : M\to M^\prime$ in $\Man$ we can assign the functor
$\DD(f) : \DD(M) \to \DD(M^\prime)$ which maps any contractible open subset $U\subseteq M$
to its image $f(U)\subseteq M^\prime$. In summary, we have defined a functor
\begin{flalign}\label{eqn:ballfunctor}
\DD : \Man \longrightarrow \mathsf{Cat}
\end{flalign}
to the category $\mathsf{Cat}$ of small categories.
\sk

Let now $M$ be any object in $\Man$. Notice that any object $U$ in $\DD(M)$ carries
a canonical orientation by pulling back the orientation of $M$ under the subset inclusion
$U\subseteq M$. Hence we may regard $\DD(M)$ as a subcategory of $\Contr$
and we can restrict the functor $\Conf : \Contr^\op \to \Ch_{\geq 0}(\Ab)$ to a functor on $\DD(M)^\op$,
which we shall denote by $\Conf_M : \DD(M)^\op \rightarrow \Ch_{\geq 0}(\Ab)$.
The functor $\Conf_M$ assigns to any contractible open subset $U\subseteq M$
the chain complex $\Conf(U)$ given by (\ref{eqn:chainYM})
and to any subset inclusion $U\subseteq V$ the restriction map
${~}\vert_U : \Conf(V) \to \Conf(U)$ given by (\ref{eqn:Confpullback}) for the subset inclusion $U\subseteq V$. 
Given now any morphism $f : M\to M^\prime$ in $\Man$, the functor (\ref{eqn:ballfunctor}) gives
a functor $\DD(f) : \DD(M)\to \DD(M^\prime)$, which defines a functor (denoted by the same symbol)
$\DD(f) : \DD(M)^\op \to \DD(M^\prime)^\op$ between the opposite categories.
Hence we obtain two functors
\begin{flalign}\label{eqn:conffunctorM}
\Conf_M : \DD(M)^\op \longrightarrow \Ch_{\geq 0}(\Ab)~,\qquad \Conf_{M^\prime}\circ \DD(f) : \DD(M)^\op \longrightarrow \Ch_{\geq 0}(\Ab)~,
\end{flalign}
with the same source and target category.
The pull-back (\ref{eqn:Confpullback}) then defines a natural transformation (denoted by the same symbol)
\begin{flalign}\label{eqn:confnat}
f^\ast : \Conf_{M^\prime}\circ \DD(f) \Longrightarrow \Conf_M~.
\end{flalign}

\subsection{Homotopy limit of canonical diagrams}
In this subsection we shall fix an arbitrary object $M$ in $\Man$
and study the homotopy limit of the canonical diagram $\Conf_M : \DD(M)^\op \to \Ch_{\geq 0}(\Ab)$
given in (\ref{eqn:conffunctorM}), which we denote by
\begin{flalign}\label{eqn:CextM}
\Conf^{\ext}(M) :=\Big(\bigoplus_{n\geq 0} \, \Conf^{\ext}(M)_n\,,~\delta\Big) := \holim(\Conf_M)~.
\end{flalign}
We use the techniques of  \cite[Section 16.8]{Dugger} and \cite{RodGon}, 
which we have summarized in detail in Appendix \ref{app:holim}.
Recall the explicit form of the functor $\Conf : \Contr^\op \to \Ch_{\geq 0}(\Ab)$ given by
(\ref{eqn:chainYM}) and (\ref{eqn:Confpullback}). In order to compute the homotopy 
limit (\ref{eqn:CextM}) of its restriction $\Conf_M : \DD(M)^\op \to \Ch_{\geq 0}(\Ab)$, 
we first take the cosimplicial replacement of this functor.
This yields a cosimplicial object in $\Ch_{\geq 0}(\Ab)$, see  (\ref{eqn:crepX}), (\ref{eqn:cofaceX}) and  (\ref{eqn:codegX})
for detailed expressions. Using the co-normalized Moore complex (\ref{eqn:dualMoorecomplex}), we assign to
this cosimplicial object in $\Ch_{\geq 0}(\Ab)$ a double chain complex in $\Ch_{\leq 0}(\Ch_{\geq 0}(\Ab))$, cf.\ (\ref{eqn:doubleX}),
which for our present functor $\Conf_M : \DD(M)^\op \to \Ch_{\geq 0}(\Ab)$ explicitly reads as
\begin{flalign}\label{eqn:doubleconf}
\xymatrix{
\ar[d]_-{\delta^{\mathrm{v}}}\prod\limits_{U}\, \Omega^1(U,\mathfrak{t})   &&
\ar[d]^-{\delta^{\mathrm{v}}}\ar[ll]_-{\delta^\mathrm{h}} \prod\limits_{U}\, C^\infty(U,\bbT)\\
\ar[d]_-{\delta^{\mathrm{v}}}\prod\limits_{U\subset V}\, \Omega^1(U,\mathfrak{t}) && \ar[d]^-{\delta^{\mathrm{v}}}\ar[ll]_-{\delta^\mathrm{h}}  \prod\limits_{U\subset V}\, C^\infty(U,\bbT)\\
\ar[d]_-{\delta^{\mathrm{v}}}\prod\limits_{U\subset V\subset W}\, \Omega^1(U,\mathfrak{t}) && \ar[d]^-{\delta^{\mathrm{v}}}\ar[ll]_-{\delta^\mathrm{h}}  \prod\limits_{U\subset V\subset W}\, C^\infty(U,\bbT)\\
\vdots&& \vdots
}
\end{flalign}
where the products are respectively over all objects $U$ in $\DD(M)$, over all 
proper subset inclusions $U\subset V$ (i.e.\ all non-identity arrows in $\DD(M)$), and 
in lower vertical degree over all $n$-fold proper subset inclusions.
The horizontal differential $\delta^{\mathrm{h}}$ is given by the product of the differentials in 
(\ref{eqn:chainYM}) and the vertical differential $\delta^{\mathrm{v}}$ is defined as the alternating sum of the co-face maps, see 
Appendix \ref{app:holim}.
The homotopy limit (\ref{eqn:CextM}) is then the truncated $\prod$-total complex 
of the double complex (\ref{eqn:doubleconf}), see (\ref{eqn:totalX}) and (\ref{eqn:holimX}).
Explicitly, we find $\Conf^{\ext}(M)_{n} =0$, for all $n\geq 2$, and
\begin{subequations}\label{eqn:CextMcomponents}
\begin{flalign}
\Conf^{\ext}(M)_1 &= \prod_{U} \, C^\infty(U,\bbT)~,\\[4pt]
\Conf^{\ext}(M)_0 & \subseteq \prod_{U} \, \Omega^1(U,\mathfrak{t}) \
\times \ \prod_{U\subset V}\, C^\infty(U,\bbT)~.
\end{flalign}
\end{subequations}
The degree $0$ component $\Conf^{\ext}(M)_0$ in (\ref{eqn:CextMcomponents}) is given by
the subgroup of all elements $\prod_{U} \, A_U \times\prod_{U\subset
  V} \, g_{(U\subset V)}$
satisfying the conditions
\begin{subequations}\label{eqn:CextMglueing}
\begin{flalign}\label{eqn:CextMglueingA}
A_V\big\vert_U^{} - A_U = g_{(U\subset V)} \,\dd g_{(U\subset V)}^{-1}~,
\end{flalign}
for all $U\subset V$, and
\begin{flalign}\label{eqn:CextMglueingg}
 g_{(V\subset W)}\big\vert_U^{} \, g_{(U\subset W)}^{-1}\,  g_{(U\subset V)} = 1\in C^\infty(U,\bbT)~,
\end{flalign}
\end{subequations}
for all $U\subset V \subset W$. The differential $\delta : \Conf^{\ext}(M)_1 \to \Conf^{\ext}(M)_0$ is explicitly 
given by
\begin{flalign}\label{eqn:CextMdifferential}
\delta\Big(\prod_U \, g_U\Big) = \prod_U\, g_U \,\dd g_U^{-1} \ \times
\ \prod_{U\subset V}\,  g_V\big\vert_{U}^{}~g_U^{-1}~.
\end{flalign}

\subsection{Deligne complex}
We shall show that the chain complex $\Conf^{\ext}(M)$, given by the homotopy limit (\ref{eqn:CextM}), 
is isomorphic to the Deligne complex for the canonical cover
$\DD(M)_0$ of $M$; see~\cite{Brylinski,Bouwknegt:2010zz,Szabo:2012hc} 
for details on the Deligne complex and Deligne cohomology. 
In the canonical cover $\DD(M)_0$, the Deligne complex reads as
\begin{flalign}\label{eqn:Delcomplex}
\Del(M) = \Big( \Del(M)_0 \oplus \Del(M)_1\,,~\delta^{\Del} \Big)~,
\end{flalign}
where 
\begin{flalign}\label{eqn:Delcomplex0}
\Del(M)_0 \subseteq \prod_{U} \, \Omega^1(U,\mathfrak{t}) \ \times \ \prod_{U,V}\, C^\infty(U\cap V,\bbT)
\end{flalign}
is the subgroup of all elements $\prod_U\, A_U \times \prod_{U,V}\, g_{UV}$ satisfying
the conditions
\begin{subequations}\label{eqn:Delglueing}
\begin{flalign}\label{eqn:DelglueingA}
A_V\big\vert_{U\cap V}^{} - A_U\big\vert_{U\cap V}^{} = g_{UV}\,\dd g_{UV}^{-1}~,
\end{flalign}
for all $U,V$, and
\begin{flalign}\label{eqn:Delglueingg}
g_{VW}\big\vert_{U\cap V\cap W}^{}~g^{-1}_{UW}\big\vert_{U\cap V\cap W}^{}~g_{UV}\big\vert_{U\cap V\cap W}^{} =1 \in C^\infty(U\cap V\cap W,\bbT)~,
\end{flalign}
\end{subequations}
for all $U,V,W$. The degree $1$ component is given by
\begin{flalign}\label{eqn:Delcomplex1}
\Del(M)_1 = \prod_{U} \, C^\infty(U,\bbT) \ ,
\end{flalign}
and the differential $\delta^{\Del} : \Del(M)_1\to \Del(M)_0$  reads as
\begin{flalign}\label{eqn:Deldifferential}
\delta^{\Del}\Big(\prod_{U}\, g_U\Big) = 
\prod_{U}\, g_U\,\dd g^{-1}_U \ \times \ \prod_{U,V}\, g_V\big\vert_{U\cap V}^{}~g_U^{-1}\big\vert^{}_{U\cap V}~.
\end{flalign}

We define a $\Ch_{\geq 0}(\Ab)$-morphism 
\begin{flalign}\label{eqn:psimap}
\psi : \Del(M)\longrightarrow \Conf^{\ext}(M)
\end{flalign}
by setting the identity $\psi_1 = \id$ on $\prod_{U}\, C^\infty(U,\bbT)$
in degree $1$ and
\begin{flalign}
\psi_0 \Big(\prod_{U}\, A_U \ \times \ \prod_{U,V} \, g_{UV}\Big) =
\prod_{U} \, A_U \ \times \ \prod_{U\subset V}\, g_{UV}
\end{flalign}
in degree $0$. Using (\ref{eqn:Delglueing}), it is easy to show that the image of $\psi_0$ lies in $\Conf^{\ext}(M)_0$,
i.e.\ that the conditions (\ref{eqn:CextMglueing}) are fulfilled.
Using also (\ref{eqn:CextMdifferential}) and (\ref{eqn:Deldifferential}) one easily shows that $\psi$
preserves the differentials, i.e.\ $\delta \circ \psi_1 = \psi_0 \circ \delta^{\Del}$.
\sk

Let us now define a $\Ch_{\geq 0}(\Ab)$-morphism
\begin{flalign}\label{eqn:varphimap}
\varphi : \Conf^{\ext}(M) \longrightarrow \Del(M)
\end{flalign}
by setting the identity $\varphi_1 = \id $ on $\prod_{U}\, C^\infty(U,\bbT)$ in degree $1$ and 
\begin{flalign}\label{eqn:varphi0tmp}
\varphi_0\Big(\prod_{U}\, A_U \ \times \ \prod_{U\subset V}\, g_{(U\subset V)} \Big) = 
\prod_{U} \, A_U \ \times \ \prod_{U,V}\, \widetilde{g}_{UV}
\end{flalign}
in degree $0$, where the functions $\widetilde{g}_{UV} \in C^\infty(U\cap V,\bbT)$ are defined by the following gluing
construction: Let us denote by $\{U_i : i\in \mathcal{I}\}$ the set of all contractible open subsets  of $M$ which are  
strictly contained in $U\cap V$. Then $\{U_i : i\in \mathcal{I}\}$ is an open cover of $U\cap V$ and we define
\begin{flalign}
(\widetilde{g}_{UV})_i := g_{(U_i\subset V)}~g_{(U_i\subset U)}^{-1} \in C^\infty(U_i,\bbT)~,
\end{flalign}
for all $i\in\mathcal{I}$. Given now $i,j\in\mathcal{I}$ such that $U_i\cap U_j\neq\emptyset$, there exists a subset $\mathcal{K}\subseteq \mathcal{I}$
such that $\{U_k : k\in\mathcal{K}\}$ is an open cover of $U_i\cap U_j$. Given any element $U_k$ of that cover, 
the conditions (\ref{eqn:CextMglueingg}) imply that
\begin{flalign}
(\widetilde{g}_{UV})_i\big\vert_{U_k}^{} = (\widetilde{g}_{UV})_k = (\widetilde{g}_{UV})_j\big\vert_{U_k}^{}~.
\end{flalign}
Hence $(\widetilde{g}_{UV})_i$ and $(\widetilde{g}_{UV})_j$ coincide on the overlap $U_i\cap U_j$.
Using now the fact that $C^\infty(\,-\,,\bbT)$ is a sheaf of Abelian groups, there exists 
an element $\widetilde{g}_{UV}\in  C^\infty(U\cap V,\bbT)$ such that  $\widetilde{g}_{UV}\vert_{U_i}^{} =(\widetilde{g}_{UV})_i $,
for all $i\in\mathcal{I}$. 
This is the element appearing on the right-hand side of (\ref{eqn:varphi0tmp}).
Using  (\ref{eqn:CextMglueing}),  it is easy to show that the image of $\varphi_0$ lies in $\Del(M)_0$,
i.e.\ that the conditions (\ref{eqn:Delglueing}) are fulfilled.
Using also (\ref{eqn:CextMdifferential}) and (\ref{eqn:Deldifferential}) one easily shows that $\varphi$
preserves the differentials.
The two $\Ch_{\geq 0}(\Ab)$-morphisms $\psi$ and $\varphi$ are inverse to each other,
which implies that $\Conf^{\ext}(M)$ and $\Del(M)$ are isomorphic.
\sk

Because the cover $\DD(M)_0$
consists of {\em contractible} open subsets of $M$, 
any principal $\bbT$-bundle connection pair on $M$ can be trivialized on this cover
and hence it can be described by an element in $\Del(M)_0$. (For this statement it does not matter
that the intersections $U\cap V$ are in general non-contractible.) Conversely,
we can construct for any element in $\Del(M)_0$ a principal $\bbT$-bundle connection pair. 
Furthermore, it is easy to check from the definition of $\delta^\Del:\Del(M)_1\to\Del(M)_0$ 
that its kernel corresponds to locally constant $\bbT$-valued
functions on $M$, namely the cohomology group $H^0(M,\bbT)$
classifying global gauge transformations. 
Using in addition the isomorphism between $\Conf^{\ext}(M)$ and $\Del(M)$, we have a chain of isomorphisms
\begin{flalign}\label{eqn:confhomologyisos}
H_\ast ( \Conf^\ext(M) ) \simeq H_\ast ( \Del(M)) \simeq \widehat{H}^2(M;\bbZ)\oplus H^0(M,\bbT)~,
\end{flalign}
where $\widehat{H}^2(M;\bbZ)$ is the second differential cohomology group, i.e.\ the Abelian group which
classifies principal $\bbT$-bundles with connection on $M$ (up to isomorphism).
In summary, we have shown that the chain complex $\Conf^\ext(M)$ given by the homotopy limit (\ref{eqn:CextM})
describes all possible global gauge field configurations on $M$. In particular, whenever $H^2(M,\bbZ)\neq 0$, 
the chain complex $\Conf^\ext(M)$ accounts for non-trivial principal $\bbT$-bundles on $M$.

\subsection{Functoriality}
We can assign to any object $M$ in $\Man$ the chain complex
$\Conf^\ext(M)$ given by the homotopy limit (\ref{eqn:CextM}).
Using  (\ref{eqn:confnat}) and functoriality of the homotopy limit it immediately follows
that this assignment is a functor
\begin{flalign}\label{eqn:Cextfunctor}
\Conf^\ext : \Man^\op \longrightarrow \Ch_{\geq 0}(\Ab)~.
\end{flalign}
Explicitly, for any morphism
$f^\op : M^\prime \to M$ in $\Man^\op$ (i.e.\ any morphism
$f : M\to M^\prime$ in $\Man$) there is a $\Ch_{\geq 0}(\Ab)$-morphism
\begin{flalign}
\Conf^\ext(f^\op) := f^\ast : \Conf^\ext(M^\prime) \longrightarrow \Conf^\ext(M)
\end{flalign}
given in degree $0$ by
\begin{subequations}
\begin{flalign}
f^\ast\Big(\prod_{U^\prime} \, A^\prime_{U^\prime} \ \times \
\prod_{U^\prime\subset V^\prime} \, g^\prime_{(U^\prime\subset V^\prime)}\Big)
=\prod_{U}\, f^\ast\big(A^\prime_{f(U)}\big) \ \times \ \prod_{U\subset V}\, f^\ast\big(g^\prime_{(f(U)\subset f(V))}\big)~,
\end{flalign}
and in degree $1$ by
\begin{flalign}
f^\ast\Big(\prod_{U^\prime}\, g^\prime_{U^\prime}\Big) = \prod_{U}\, f^\ast\big(g^\prime_{f(U)}\big)~.
\end{flalign}
\end{subequations}

\subsection{Functor extension}
We shall show that the functor $\Conf^\ext : \Man^\op \to \Ch_{\geq 0}(\Ab)$ given in
(\ref{eqn:Cextfunctor}) is an extension of our original functor $\Conf : \Contr^\op \to \Ch_{\geq 0} (\Ab)$, 
i.e.\ that there is a diagram of functors
\begin{flalign}\label{eqn:confextensiondiagram}
\xymatrix{
\ar[dr]\Contr^\op \ar[rr]^-{\Conf} &\ar@{=>}[d]^-{\eta} & \Ch_{\geq 0}(\Ab)\\
& \Man^\op \ar[ru]_-{\Conf^\ext}&
}
\end{flalign}
which commutes up to a natural transformation $\eta$. The functor $\Contr^\op \to \Man^\op$ 
is simply the full subcategory embedding. We further show that the natural transformation $\eta$ 
is a natural quasi-isomorphism, so that the functors $\Conf$ and $\Conf^\ext$ 
give weakly equivalent descriptions of the gauge field configurations on contractible manifolds.
Our extension $\Conf^\ext$ of $\Conf$ is distinguished by the fact that 
it gives a correct description of the global gauge field configurations on non-contractible manifolds, see
(\ref{eqn:confhomologyisos}).
\sk

For any object $M$ in $\Contr$, there is a $\Ch_{\geq 0}(\Ab)$-morphism
\begin{flalign}\label{eqn:etaM}
\eta_M : \Conf(M)\longrightarrow \Conf^\ext(M)
\end{flalign}
given by
\begin{flalign}
{\eta_M}_0(A) = \prod_{U} \, A\vert_{U}^{} \ \times \ \prod_{U\subset
  V} \, 1~,\qquad
{\eta_M}_1(g) = \prod_{U}\, g\big\vert_{U}^{}~.
\end{flalign}
One easily checks that $\eta_M$ are the components of a natural transformation, i.e.\ for all
morphisms $f: M\to M^\prime$ in $\Contr$ there is a commutative diagram
\begin{flalign}
\xymatrix{
\ar[d]_-{f^\ast}\Conf(M^\prime) \ar[rr]^-{\eta_{M^\prime}} && \Conf^{\ext}(M^\prime)\ar[d]^-{f^\ast}\\
\Conf(M) \ar[rr]_-{\eta_{M}} && \Conf^\ext(M)
}
\end{flalign}
in the category $\Ch_{\geq 0}(\Ab)$. Hence the diagram of functors (\ref{eqn:confextensiondiagram}) 
commutes up to the natural transformation $\eta$ and $\Conf^\ext$ is an extension of $\Conf$.
\sk

It remains to show that $\eta$ is a natural quasi-isomorphism, i.e.\ that any component
$\eta_M$ is a quasi-isomorphism in $\Ch_{\geq 0}(\Ab)$.  For this, we define a (non-natural)
$\Ch_{\geq 0}(\Ab)$-morphism
\begin{flalign}
\theta_M : \Conf^\ext(M)\longrightarrow \Conf(M)
\end{flalign}
by setting
\begin{flalign}
{\theta_M}_0\Big(\prod_{U}\, A_U \ \times \ \prod_{U\subset V}\, g_{(U\subset V)}\Big) = A_M~,\qquad
{\theta_M}_1\Big(\prod_{U}\, g_U\Big) = g_M~.
\end{flalign}
Notice that $\theta_M\circ\eta_M = \id$ and that $\eta_M\circ \theta_M -\id = \delta\circ h + h\circ \delta$
with chain homotopy
\begin{flalign}
h : \Conf^\ext(M)_0\longrightarrow \Conf^\ext(M)_1~,~~\prod_{U}\, A_U
\ \times \ \prod_{U\subset V}\, g_{(U\subset V)}\longmapsto
\prod_{U}\, g_{(U\subset M)}~.
\end{flalign}
Hence any $\Ch_{\geq 0}(\Ab)$-morphism (\ref{eqn:etaM}) is a quasi-isomorphism.


\section{\label{sec:globalobservables}Homotopy colimits and global gauge field observables}
Our functor $\Obs : \Contr \to \Ch_{\leq 0}(\Ab)$ given by (\ref{eqn:obschainYM}) and (\ref{eqn:Obspushforward})
describes the chain complexes of gauge field observables (given by smooth group characters)
of Abelian gauge theory on {\em contractible}  manifolds. For an arbitrary
 manifold $M$, the chain complex in (\ref{eqn:obschainYM}) does not describe sufficiently
many observables to separate all gauge field configurations on $M$. In particular,
if $H^2(M,\bbZ)\neq 0$, there are non-trivial principal $\bbT$-bundles
over $M$ which are not measured by the observables in (\ref{eqn:obschainYM}).
In this section we perform the dual of the construction in Section \ref{sec:globalconf}
in order to extend the functor $\Obs : \Contr \to \Ch_{\leq 0}(\Ab)$ to the category $\Man$
of all oriented $m$-dimensional manifolds.

\subsection{Canonical diagrams of gauge field observables}
Recalling the functor $\DD : \Man \to \mathsf{Cat}$ given in Subsection \ref{subsec:canonicalcover}
(cf.\ (\ref{eqn:ballfunctor})) and the fact that $\DD(M)$ may be regarded as a subcategory of $\Contr$, 
for any object $M$ in $\Man$,
we can restrict the functor $\Obs : \Contr\to \Ch_{\leq 0}(\Ab)$ to a functor
on $\DD(M)$, which we shall denote by $\Obs_M : \DD(M)\rightarrow \Ch_{\leq 0}(\Ab)$.
The functor $\Obs_M$ assigns to any contractible open subset $U\subseteq M$ the chain complex
$\Obs(U)$ given by (\ref{eqn:obschainYM}) and to any subset inclusion
$U\subseteq V$ the extension (by zero) map $\ext_{V} : \Obs(U)\to\Obs(V)$
given by (\ref{eqn:Obspushforward}) for the subset inclusion $U\subseteq V$. 
Given now any morphism $f : M\to M^\prime$ in $\Man$, the functor (\ref{eqn:ballfunctor})
gives a functor $\DD(f) : \DD(M)\to \DD(M^\prime)$, hence there are two parallel functors
\begin{flalign}\label{eqn:obsfunctorM}
\Obs_M : \DD(M)\longrightarrow \Ch_{\leq 0}(\Ab)~,\qquad
\Obs_{M^\prime}\circ \DD(f) : \DD(M)\longrightarrow \Ch_{\leq 0}(\Ab)~.
\end{flalign}
The push-forward (\ref{eqn:Obspushforward}) then defines a natural transformation (denoted by the same symbol)
\begin{flalign}\label{eqn:obsnat}
f_\ast : \Obs_{M} \Longrightarrow \Obs_{M^\prime}\circ \DD(f)~.
\end{flalign}

\subsection{Homotopy colimit of canonical diagrams}
We fix any object $M$ in $\Man$ and study the homotopy colimit of the canonical
diagram $\Obs_M :\DD(M)\to \Ch_{\leq 0}(\Ab)$ given in (\ref{eqn:obsfunctorM}),
which we denote by
\begin{flalign}\label{eqn:ObsextM}
\Obs^{\ext}(M) := \Big(\bigoplus_{n\leq 0} \, \Obs^{\ext}(M)_n\,,~\delta^\ast\Big)
:= \hocolim(\Obs_M)~. 
\end{flalign}
We use the techniques summarized in Appendix \ref{app:hocolim}. 
Recall the explicit form of the functor $\Obs : \Contr \to \Ch_{\leq 0}(\Ab)$
given by (\ref{eqn:obschainYM}) and (\ref{eqn:Obspushforward}).
In order to compute the homotopy 
colimit (\ref{eqn:ObsextM}) of its restriction $\Obs_M : \DD(M) \to \Ch_{\leq 0}(\Ab)$, 
we first take the simplicial replacement of this functor.
This yields a simplicial object in $\Ch_{\leq 0}(\Ab)$, see (\ref{eqn:srepY}), (\ref{eqn:faceY}) and (\ref{eqn:degY})
for detailed expressions. Using the normalized Moore complex (\ref{eqn:Moorecomplex}), we assign to
this simplicial object in $\Ch_{\leq 0}(\Ab)$ a double chain complex in $\Ch_{\geq 0}(\Ch_{\leq 0}(\Ab))$, cf.\ (\ref{eqn:doubleY}),
which for our present functor $\Obs_M : \DD(M) \to \Ch_{\leq 0}(\Ab)$ explicitly reads as
\begin{flalign}\label{eqn:doubleobs}
\xymatrix{
\ar[d]_-{\delta^{\mathrm{v}}}\vdots &&\vdots\ar[d]^-{\delta^{\mathrm{v}}}\\
\ar[d]_-{\delta^{\mathrm{v}}}\coprod\limits_{U\subset V\subset W}\,\Omega^m_{\cc,\bbZ}(U) &&\ar[d]^-{\delta^{\mathrm{v}}} \coprod\limits_{U\subset V\subset W}\,\Omega^{m-1}_\cc(U)\ar[ll]_-{\delta^\mathrm{h}}\\
\ar[d]_-{\delta^{\mathrm{v}}}\coprod\limits_{U\subset V}\,\Omega^m_{\cc,\bbZ}(U) &&\ar[d]^-{\delta^{\mathrm{v}}} \coprod\limits_{U\subset V}\,\Omega^{m-1}_\cc(U)\ar[ll]_-{\delta^\mathrm{h}}\\
\coprod\limits_{U}\,\Omega^m_{\cc,\bbZ}(U) && \coprod\limits_{U}\,\Omega^{m-1}_\cc(U)\ar[ll]_-{\delta^\mathrm{h}}
}
\end{flalign}
where the coproducts are respectively over all objects $U$ in $\DD(M)$, over all 
proper subset inclusions $U\subset V$, and 
in higher vertical degree over all $n$-fold proper subset inclusions.
The horizontal differential $\delta^{\mathrm{h}}$ is given by the coproduct of the differentials in 
(\ref{eqn:obschainYM}) and the vertical differential $\delta^{\mathrm{v}}$ is defined as the alternating sum of the face maps, 
see Appendix \ref{app:hocolim}.
The homotopy colimit (\ref{eqn:ObsextM}) is then the truncated $\coprod$-total complex 
of the double complex (\ref{eqn:doubleobs}), see (\ref{eqn:totalY}) and (\ref{eqn:hocolimY}).
Explicitly, we find $\Obs^\ext(M)_n = 0$, for all $n\leq -2$, and
\begin{subequations}\label{eqn:ObsextMcomponents}
\begin{flalign}
\Obs^\ext(M)_{-1} &= \coprod_{U} \, \Omega^m_{\cc,\bbZ}(U)~,\\[4pt]
\Obs^\ext(M)_0 &= \Big(\coprod_{U}\, \Omega^{m-1}_{\cc}(U) \ \oplus \
\coprod_{U\subset V}\, \Omega^m_{\cc,\bbZ}(U)\Big) \, \Big/ \,
\mathfrak{I}(M)~.
\end{flalign}
\end{subequations}
The quotient in $\Obs^\ext(M)_0$ is by the Abelian subgroup $\mathfrak{I}(M)$ that is generated by the elements
\begin{subequations}\label{eqn:ObsextMrelations}
\begin{flalign}
\iota_U(\varphi) - \iota_V\big(\ext_V(\varphi)\big) - \iota_{(U\subset V)}(\dd \varphi)~,
\end{flalign}
for all $U\subset V$ and $\varphi\in \Omega^{m-1}_{\cc}(U)$, and
\begin{flalign}
\iota_{(U\subset V)}(\chi) - \iota_{(U\subset W)}(\chi) + \iota_{(V\subset W)}\big(\ext_V(\chi)\big)~,
\end{flalign}
\end{subequations}
for all $U\subset V\subset W$ and $\chi\in \Omega^{m}_{\cc,\bbZ}(U)$.
Here  $\iota_{-}$ denote the inclusion morphisms in the coproducts and
as before $\ext_{-}$ denote the
extension maps.
 The differential $\delta^\ast : \Obs^\ext(M)_0\to \Obs^\ext(M)_{-1}$ is explicitly given by
 \begin{flalign}\label{eqn:ObsextMdifferential}
 \delta^\ast\big(\iota_U(\varphi)\big) = \iota_U(\dd\varphi)~,\qquad \delta^\ast\big(\iota_{(U\subset V)}(\chi)\big)
 =\iota_{U}(\chi) - \iota_V\big(\ext_V(\chi)\big)~,
 \end{flalign}
 where we suppress the equivalence classes in $\Obs^\ext(M)_0$.

\subsection{Functoriality}
We can assign to any object $M$ in $\Man$ the chain complex $\Obs^\ext(M)$ given by the
homotopy colimit (\ref{eqn:ObsextM}). Using (\ref{eqn:obsnat}) and functoriality of the homotopy colimit, it immediately follows
that this assignment is a functor
\begin{flalign}\label{eqn:Oextfunctor}
\Obs^\ext : \Man \longrightarrow \Ch_{\leq 0}(\Ab)~.
\end{flalign}
Explicitly, for any morphism $f : M \to M^\prime$ in $\Man$ there is
a $\Ch_{\leq 0}(\Ab)$-morphism
\begin{flalign}
\Obs^\ext(f) := f_\ast : \Obs^\ext(M)\longrightarrow \Obs^\ext(M^\prime)
\end{flalign}
given in degree $0$ by
\begin{subequations}
\begin{flalign}
f_\ast\big(\iota_U(\varphi)\big) = \iota_{f(U)}\big(f_\ast(\varphi)\big)~,\qquad
f_\ast\big(\iota_{(U\subset V)}(\chi)\big)  = \iota_{(f(U)\subset f(V))}\big(f_\ast(\chi)\big)~,
\end{flalign}
and in degree $-1$ by
\begin{flalign}
f_\ast\big(\iota_U(\chi)\big) = \iota_{f(U)}\big(f_\ast(\chi)\big)~.
\end{flalign}
\end{subequations}

\subsection{Functor extension}
We shall show that the functor $\Obs^\ext : \Man\to \Ch_{\leq 0}(\Ab)$ given in (\ref{eqn:Oextfunctor})
is an extension of our original functor $\Obs : \Contr \to \Ch_{\leq 0}(\Ab)$, i.e.\ that there is a
diagram of functors
\begin{flalign}
\xymatrix{
\ar[rd]\Contr \ar[rr]^-{\Obs} && \Ch_{\leq 0}(\Ab)\\
&\Man\ar[ru]_-{\Obs^\ext}\ar@{=>}[u]_-{\zeta}&
}
\end{flalign}
which commutes up to a natural transformation $\zeta$.
We further show that $\zeta$ is a natural quasi-isomorphism,
so that the functors $\Obs$ and $\Obs^\ext$ give weakly equivalent
descriptions of the gauge field observables on contractible manifolds.
\sk

For any object $M$ in $\Contr$, there is a $\Ch_{\leq 0}(\Ab)$-morphism
\begin{flalign}\label{eqn:zetamap}
\zeta_{M} : \Obs^\ext(M) \longrightarrow \Obs(M)
\end{flalign}
given in degree $0$ by
\begin{subequations}
\begin{flalign}
{\zeta_M}_0\big(\iota_U(\varphi)\big) = \ext_M(\varphi)~,
\qquad {\zeta_M}_0\big(\iota_{(U\subset V)}(\chi)\big) = 0~,
\end{flalign}
and in degree $-1$ by
\begin{flalign}
{\zeta_M}_{-1}\big(\iota_U(\chi)\big) = \ext_M(\chi)~.
\end{flalign}
\end{subequations}
One easily checks that $\zeta_M$ are the components of a natural transformation,
i.e.\ for all morphisms $f : M\to M^\prime$ in $\Contr$
there is a commutative diagram
\begin{flalign}
\xymatrix{
\ar[d]_-{f_\ast}\Obs^\ext(M) \ar[rr]^-{\zeta_M} && \Obs(M)\ar[d]^-{f_\ast}\\
\Obs^\ext(M^\prime) \ar[rr]_-{\zeta_{M^\prime}}&&\Obs(M^\prime)
}
\end{flalign}
in the category $\Ch_{\leq 0}(\Ab)$.
\sk

It remains to show that $\zeta$ is a natural quasi-isomorphism, i.e.\ that any component
$\zeta_M$ is a quasi-isomorphism in $\Ch_{\leq 0}(\Ab)$.
For this, we define a (non-natural) $\Ch_{\leq 0}(\Ab)$-morphism
\begin{flalign}
\kappa_M : \Obs(M)\longrightarrow \Obs^\ext(M)
\end{flalign}
by setting 
\begin{flalign}
{\kappa_M}_0(\varphi) = \iota_M(\varphi)~,\qquad
{\kappa_M}_{-1}(\chi) = \iota_M(\chi)~.
\end{flalign}
Notice that $\zeta_M \circ \kappa_M = \id$
and that $\kappa_M\circ \zeta_M -\id = \delta^\ast \circ k + k\circ \delta^\ast$
with chain homotopy
\begin{flalign}
k :  \Obs^\ext(M)_{-1}\longrightarrow \Obs^\ext(M)_0~,~~\iota_U(\chi) \longmapsto -\iota_{(U\subset M)}(\chi)~.
\end{flalign}
Hence any $\Ch_{\leq 0}(\Ab)$-morphism (\ref{eqn:zetamap}) is a quasi-isomorphism.

\subsection{Natural pairing}
For any object $M$ in $\Man$, there is a grading-preserving pairing given by the bi-character
\begin{flalign}\label{eqn:pairingext}
\ip{-}{-}_M^\ext : \Obs^\ext(M)\times \Conf^\ext(M)\longrightarrow \bbT
\end{flalign}
defined by
\begin{subequations}
\begin{flalign}
\Big\langle \iota_V(\chi) \,,\, \prod_{U}g_U\, \Big\rangle_M^\ext &:= \ip{\chi}{g_V}_V~,\\[4pt]
\Big\langle\iota_W(\varphi) \,,\, \prod_{U}\, A_U \ \times \
\prod_{U\subset V} \, g_{(U\subset V)}\Big\rangle_M^\ext & := \ip{\varphi}{A_W}_W~,\\[4pt]
\Big\langle\iota_{(W\subset X)}(\chi) \,,\, \prod_{U}\, A_U \ \times \
\prod_{U\subset V}\, g_{(U\subset V)}\Big\rangle_{M}^\ext & :=
\ip{\chi}{g^{-1}_{(W\subset X)}}_{W}~,  
\end{flalign}
\end{subequations}
where the right-hand sides are given by the pairings (\ref{eqn:pairing}) for contractible manifolds. 
Using the conditions (\ref{eqn:CextMglueing}), one immediately checks that this pairing 
is compatible with the quotient in $\Obs^\ext(M)_0$
that is generated by the elements (\ref{eqn:ObsextMrelations}).
Moreover, the differentials $\delta$ in $\Conf^\ext(M)$ and $\delta^\ast$ in $\Obs^\ext(M)$ 
are dual to each other via the pairing (\ref{eqn:pairingext}), i.e.\
\begin{flalign}
\ip{\delta^\ast F}{B}^\ext_M = \ip{F}{\delta B}^\ext_M~,
\end{flalign}
for all $F\in \Obs^\ext(M)$ and $B\in \Conf^\ext(M)$.
The pairing (\ref{eqn:pairingext}) is natural in the sense that
the diagram
\begin{flalign}
\xymatrix{
\ar[d]_-{f_\ast\times \id}\Obs^\ext(M) \times \Conf^\ext(M^\prime) \ar[rr]^-{\id \times f^\ast} && \Obs^\ext(M)\times \Conf^\ext(M)\ar[d]^-{\ips{-}{-}^\ext_M}\\
\Obs^\ext(M^\prime)\times \Conf^\ext(M^\prime)\ar[rr]_-{\ips{-}{-}^\ext_{M^\prime}}&&\bbT
}
\end{flalign}
 commutes for all morphisms $f: M\to M^\prime$ in $\Man$.
\sk

Notice that the pairing (\ref{eqn:pairingext}) is non-degenerate in the right entry, i.e.\ 
the observables $\Obs^\ext(M)$ separate all possible global gauge field 
configurations $\Conf^\ext(M)$ on $M$. In particular, when
$H^2(M,\bbZ)\neq 0$, our homotopy colimit construction has produced enough observables 
to measure and distinguish all possible principal $\bbT$-bundles on $M$.



\section*{Acknowledgements}
We thank Andr{\'e} Henriques and especially Urs Schreiber 
for many helpful comments on the material presented in this paper. We also thank an anonymous referee for prompting us to clarify some details of our constructions. This work was completed while
R.J.S. was visiting the Alfred Renyi Institute of Mathematics
in Budapest during March--April 2015, whom he warmly thanks for support and hospitality during his
stay there.
The work of M.B.\ is supported by a Research Fellowship of the Della
Riccia Foundation (Italy).
The work of A.S.\ is supported by a Research Fellowship of the Deutsche
Forschungsgemeinschaft (DFG, Germany). 
The work of R.J.S.\ is partially supported by the Consolidated Grant
ST/L000334/1 from the
UK Science and Technology Facilities Council, and by the Advanced Grant LDTBUD from the European Research Council.


\appendix

\section{\label{app:DoldKan}Dold-Kan correspondence and Moore complex}
We shall briefly review the Dold-Kan correspondence 
between simplicial Abelian groups and non-negatively graded 
chain complexes of Abelian groups. In particular, we shall give explicit
formulas for the normalized Moore complex, which is used at various instances
in this paper. For further details and full proofs, see~\cite[Section III.2]{GJ99}.
\sk

Denoting by $\mathsf{sAb}$ the category of simplicial Abelian groups and by $\Ch_{\geq 0}(\Ab)$
the category of non-negatively graded chain complexes of Abelian groups, 
the {\em Dold-Kan correspondence}  states that there are two functors
\begin{flalign}
N : \mathsf{sAb}\longrightarrow \Ch_{\geq 0}(\Ab) \ , \qquad \Gamma : \Ch_{\geq 0}(\Ab)\longrightarrow \mathsf{sAb}~,
\end{flalign}
which form an equivalence of categories, see \cite[Section III.2, Corollary 2.3]{GJ99}.
For the purposes of the present paper, we only need an explicit description of the functor
$N$, which is called the {\em normalized Moore complex}.
Let $A=\{A_n\}_{n\in\bbN_0}$ be any simplicial Abelian group with face and degeneracy maps
denoted by $\partial_i^n : A_{n}\to A_{n-1}$, for $i=0,1,\dots,n$ and $n\geq 1$, 
and $\epsilon_{i}^n : A_{n}\to A_{n+1}$, for $i=0,1,\dots,n$ and $n\geq 0$.
 Then the functor $N$ assigns to $A$ the non-negatively graded chain complex of Abelian groups
 \begin{subequations}\label{eqn:Moorecomplex}
 \begin{flalign}
N(A) := \Big(\bigoplus_{n\geq 0} \, N(A)_n\,,~\delta\Big)~,
\end{flalign}
where
\begin{flalign}
N(A)_n := \frac{A_n}{\epsilon^{n-1}_0(A_{n-1}) + \cdots
  +\epsilon^{n-1}_{n-1}(A_{n-1}) } \ ,
\end{flalign}
for all $n\geq 0$, and the differential $\delta$ (of degree $-1$) is defined as the alternating
sum of the face maps, i.e.\ we set
\begin{flalign}
\delta := \sum_{i=0}^n \, (-1)^{i}\ \partial_i^n
\end{flalign}
\end{subequations}
on  $N(A)_n$.
\sk

The {\em dual Dold-Kan correspondence} is an equivalence between the categories
of cosimplicial Abelian groups $\mathsf{cAb}$ and non-positively graded chain 
complexes of Abelian groups $\Ch_{\leq 0}(\Ab)$. For our purposes we only have to explain the
functor $N^\ast : \mathsf{cAb} \to \Ch_{\leq 0}(\Ab)$, which is called the {\em co-normalized Moore complex}.
Let $A=\{A_n\}_{n\in\bbN_0}$ be any cosimplicial Abelian group with co-face and co-degeneracy maps
denoted by $d^i_n : A_{n}\to A_{n+1}$, for $i=0,1,\dots,n+1$ and $n\geq 0$, 
and $e^{i}_n : A_{n}\to A_{n-1}$, for $i=0,1,\dots,n-1$ and $n\geq 1$.
 Then the functor $N^\ast$ assigns to $A$ the non-positively graded chain complex of Abelian groups
 \begin{subequations}\label{eqn:dualMoorecomplex}
 \begin{flalign}
N^\ast(A) := \Big(\bigoplus_{n\leq 0} \, N^\ast(A)_n\,,~\delta^\ast\Big)~,
\end{flalign}
where
\begin{flalign}
N^\ast(A)_{-n} := \bigcap_{i=0}^{n-1} \, \Ker\big(e_n^i : A_n\to
A_{n-1}\big) \ ,
\end{flalign}
for all $n\geq 0$, and the differential $\delta^\ast$ (of degree $-1$) is defined as the alternating
sum of the co-face maps, i.e.\ we set
\begin{flalign}
\delta^\ast := \sum_{i=0}^{n+1}\, (-1)^{i}\ d^i_n
\end{flalign}
\end{subequations}
on  $N^\ast(A)_{-n}$.
\sk

Note that the normalized Moore complex
(\ref{eqn:Moorecomplex}) is still defined when we
replace the category of Abelian groups $\Ab$ by other Abelian categories,
such as the category of (possibly unbounded) chain complexes of Abelian groups $\Ch(\Ab)$.
In this case the normalized Moore complex $N$ assigns to simplicial chain complexes of Abelian groups
$\mathsf{s}\Ch(\Ab)$ double chain complexes of Abelian groups $\Ch_{\geq 0}(\Ch(\Ab))$, where the first grading
is non-negative. Similarly, the co-normalized Moore complex (\ref{eqn:dualMoorecomplex})
is still defined when we replace the category of  Abelian groups $\Ab$ by $\Ch(\Ab)$.
Then the co-normalized Moore complex $N^\ast$ assigns to cosimplicial chain complexes
of Abelian groups $\mathsf{c}\Ch(\Ab)$ double chain complexes of Abelian groups $\Ch_{\leq 0}(\Ch(\Ab))$, 
where the first grading is non-positive.


\section{\label{app:holimcolim}Homotopy limits and colimits for chain complexes}
We shall briefly explain how to compute homotopy limits and colimits of diagrams of chain complexes of Abelian groups.
Our presentation follows mainly \cite[Section 16.8]{Dugger}, but we also refer the reader to \cite{RodGon}
for more technical details. 

\subsection{\label{app:holim}Homotopy limits for non-negatively graded chain complexes}
Let $\DD$ be a small category. Given any functor $\XX : \DD \to\Ch_{\geq 0}(\Ab)$, which we interpret as a diagram
in $\Ch_{\geq 0}(\Ab)$ of shape $\DD$, we would like to compute the
homotopy limit $\holim(\XX)$. This construction is a three step procedure:  First, one takes the cosimplicial replacement
of the diagram $\XX : \DD \to\Ch_{\geq 0}(\Ab)$, which gives a cosimplicial object in $\Ch_{\geq 0}(\Ab)$.
Then one assigns a 
double chain complex in $\Ch_{\leq 0}(\Ch_{\geq 0}(\Ab))$ via the co-normalized Moore complex, where the first grading is non-positive and the second
grading is non-negative. Finally one forms the $\prod$-total complex, 
which gives the homotopy limit $\holim(\XX)$ after truncation to non-negative degrees. 
We shall now explain these steps in more detail and give explicit formulas.
\sk

The nerve of the small category $\DD$ is the simplicial set $\{\DD_n\}_{n\in\bbN_0}$,
where $\DD_0$ is the set of objects in $\DD$ and $\DD_n$, for $n\geq 1$,  is the set
of all composable $n$-arrows in $\DD$. For $n\geq 1$, we shall denote an element of $\DD_n$
by an $n$-tuple $(f_1,\dots,f_n)$ of morphisms in $\DD$ such that
the source of $f_i$ is the target of $f_{i+1}$ (i.e.\ the compositions $f_i\circ f_{i+1}$ exist).
The face maps are given by composing two subsequent arrows (or throwing away the first/last arrow)
and the degeneracy maps are given by inserting the identity morphisms.
The cosimplicial replacement of $\XX :\DD \to \Ch_{\geq 0}(\Ab)$ is the cosimplicial object
in $\Ch_{\geq 0}(\Ab)$ given by
\begin{flalign}\label{eqn:crepX}
\xymatrix{
\prod\limits_{d\in \DD_0}\, \XX(d) \ar@<0.5ex>[r]\ar@<-0.5ex>[r] & 
\prod\limits_{f\in \DD_1}\, \XX(\mathrm{t}(f)) 
\ar@<1ex>[r]\ar@<-1ex>[r]\ar[r] &  \prod\limits_{(f_1,f_2)\in \DD_2}\, \XX(\mathrm{t}(f_1)) 
\ar@<0.5ex>[r]\ar@<-0.5ex>[r] \ar@<1.5ex>[r]\ar@<-1.5ex>[r] &~\cdots
}~,
\end{flalign}
where the arrows are the co-face maps and we have suppressed the co-degeneracy maps for notational convenience. 
Here $\prod$ denotes the product in the category $\Ch_{\geq 0}(\Ab)$ and we have denoted by
$\mathrm{t}(f)$ the target of a morphism $f$ in $\DD$.
The co-face maps $d_n^i : \prod_{(f_1,\dots,f_n)\in \DD_n}\, \XX(\mathrm{t}(f_1))  
\to  \prod_{(f_1,\dots,f_{n+1})\in \DD_{n+1}}\, \XX(\mathrm{t}(f_1))  $,
for $n\geq 0$ and $i=0,1,\dots,n+1$, are defined by using the universal property of the product; explicitly,
for $i=0$,
\begin{subequations}\label{eqn:cofaceX}
\begin{flalign}
\xymatrix{
\ar[d]_-{\pi_{(h_2,\dots ,h_{n+1})}}\prod\limits_{(f_1,\dots,f_n)\in \DD_n}\, \XX(\mathrm{t}(f_1)) \ar@{.>}[rr]^{d_n^0}  && \ar[d]^-{\pi_{(h_1,\dots ,h_{n+1})}}\prod\limits_{(f_1,\dots,f_{n+1})\in \DD_{n+1}}\, \XX(\mathrm{t}(f_1)) \\
 \XX(\mathrm{t}(h_2)) \ar[rr]_-{\XX(h_1)}&& \XX(\mathrm{t}(h_1)) 
}
\end{flalign}
for $1\leq i\leq n$,
\begin{flalign}
\xymatrix{
\ar[dr]_-{\pi_{(h_1,\dots, h_i\circ h_{i+1},\dots, h_{n+1})}~~~~~~~~~}\prod\limits_{(f_1,\dots,f_n)\in \DD_n}\, \XX(\mathrm{t}(f_1)) \ar@{.>}[rr]^{d_n^i}  && \ar[dl]^-{~~~~~~~\pi_{(h_1,\dots ,h_{n+1})}}\prod\limits_{(f_1,\dots,f_{n+1})\in \DD_{n+1}}\, \XX(\mathrm{t}(f_1)) \\
& \XX(\mathrm{t}(h_1)) &
}
\end{flalign}
and for $i=n+1$,
\begin{flalign}
\xymatrix{
\ar[dr]_-{\pi_{(h_1,\dots, h_{n})}~~~~~~~}\prod\limits_{(f_1,\dots,f_n)\in \DD_n}\, \XX(\mathrm{t}(f_1)) \ar@{.>}[rr]^{d_n^{n+1}}  && \ar[dl]^-{~~~~~~~\pi_{(h_1,\dots ,h_{n+1})}}\prod\limits_{(f_1,\dots,f_{n+1})\in \DD_{n+1}}\, \XX(\mathrm{t}(f_1)) \\
& \XX(\mathrm{t}(h_1)) &
}
\end{flalign}
\end{subequations}
where $\pi_-$ are the projection morphisms from the products.
The co-degeneracy maps $e_n^i  : \prod_{(f_1,\dots,f_n)\in \DD_n}\, \XX(\mathrm{t}(f_1))  
\to  \prod_{(f_1,\dots,f_{n-1})\in \DD_{n-1}}\, \XX(\mathrm{t}(f_1))  $,
for $n\geq 1$ and $i=0,1,\dots,n-1$, are also defined by using the universal property of the product; explicitly, for $i=0,1,\dots,n-1$,
\begin{flalign}\label{eqn:codegX}
\xymatrix{
\ar[dr]_-{\pi_{(h_1,\dots,h_i,\id,h_{i+1},\dots,h_{n-1})}~~~~~~~~~}\prod\limits_{(f_1,\dots,f_n)\in \DD_n}\, \XX(\mathrm{t}(f_1))\ar@{.>}[rr]^-{e_n^i}  && \prod\limits_{(f_1,\dots,f_{n-1})\in \DD_{n-1}}\, \XX(\mathrm{t}(f_1)) \ar[dl]^-{~~~~~~\pi_{(h_1,\dots,h_{n-1})}}\\
&\XX(\mathrm{t}(h_1))&
}
\end{flalign}

Using the co-normalized Moore complex (\ref{eqn:dualMoorecomplex}), we can assign to the cosimplicial object (\ref{eqn:crepX})
in $\Ch_{\geq 0}(\Ab)$ a double chain complex in $\Ch_{\leq 0}(\Ch_{\geq 0}(\Ab))$.
Denoting this double chain complex by $\XX_{\ast,\ast}$, a simple calculation shows that
\begin{flalign}\label{eqn:chainchainX}
\XX_{0,\ast} = \prod\limits_{d\in \DD_0}\, \XX(d)~,\qquad
\XX_{-n,\ast} = 
\prod\limits_{\mycom{(f_1,\dots,f_n)\in \DD_n}{f_i\neq \id }}\, \XX(\mathrm{t}(f_1))~,
\end{flalign}
for all $n\geq 1$, where the second product is taken over all composable $n$-arrows $(f_1,\dots,f_n)$
such that none of the $f_i$ is an identity morphism.
The vertical differential $\delta^\mathrm{v}  : \XX_{\ast,\ast} \to \XX_{\ast-1,\ast}$
is given by the alternating sum of the co-face maps, i.e.\ $\delta^\mathrm{v} = \sum_{i=0}^{n+1}\, (-1)^i~d_n^i$
on $\XX_{-n,\ast}$, and the horizontal differential $\delta^\mathrm{h}  : 
\XX_{\ast,\ast} \to\XX_{\ast,\ast-1}$ is given by the product of
the differentials in the chain complexes $\XX(d)$, for $d$ an object in $\DD$.
The double complex $\XX_{\ast,\ast}$ may be visualized as 
\begin{flalign}\label{eqn:doubleX}
\xymatrix{
\ar[d]_-{\delta^{\mathrm{v}}} \XX_{0,0} && \ar[d]^-{\delta^{\mathrm{v}}} \ar[ll]_-{\delta^{\mathrm{h}}} \XX_{0,1} && \ar[d]^-{\delta^{\mathrm{v}}} \ar[ll]_-{\delta^{\mathrm{h}}} \XX_{0,2} && \ar[ll]_-{\delta^{\mathrm{h}}} \cdots \\
\ar[d]_-{\delta^{\mathrm{v}}} \XX_{-1,0} && \ar[d]^-{\delta^{\mathrm{v}}} \ar[ll]_-{\delta^{\mathrm{h}}} \XX_{-1,1} && \ar[d]^-{\delta^{\mathrm{v}}} \ar[ll]_-{\delta^{\mathrm{h}}} \XX_{-1,2} && \ar[ll]_-{\delta^{\mathrm{h}}} \cdots\\
\XX_{-2,0} \ar[d]_-{\delta^{\mathrm{v}}} && \ar[d]^-{\delta^{\mathrm{v}}} \ar[ll]_-{\delta^{\mathrm{h}}}\XX_{-2,1} && \ar[d]^-{\delta^{\mathrm{v}}} \ar[ll]_-{\delta^{\mathrm{h}}}\XX_{-2,2} && \ar[ll]_-{\delta^{\mathrm{h}}} \cdots\\
\vdots&&\vdots && \vdots &&\ddots
}
\end{flalign}

We now form the $\prod$-total complex
\begin{flalign}\label{eqn:totalX}
\XX^{\mathrm{Tot}} :=\Big(\bigoplus_{n\in\bbZ}\, \XX^{\mathrm{Tot}}_n\,,\delta^{\mathrm{Tot}}\Big) 
:=  \Big(\bigoplus_{n\in\bbZ} \ \prod_{p+q=n}\, \XX_{p,q}\,,~\delta^{\mathrm{Tot}} := \delta^{\mathrm{v}} + (-1)^p\,\delta^{\mathrm{h}} \Big)
\end{flalign}
and we notice that $\XX^{\mathrm{Tot}}$ is a {\em $\bbZ$-graded} chain complex of Abelian groups, in particular
it is non-trivial in negative degrees.
The homotopy limit $\holim(\XX)$ of the diagram $\XX :\DD\to\Ch_{\geq 0}(\Ab)$ is then the truncation
of $\XX^{\mathrm{Tot}}$ to non-negative degrees. Explicitly,
\begin{subequations}\label{eqn:holimX}
\begin{flalign}
\holim(\XX) = \Big(\bigoplus_{n\geq 0} \, \holim(\XX)_n\,,~\delta\Big) ~,
\end{flalign}
where
\begin{flalign}
\holim(\XX)_0 = \Ker\big(\delta^{\mathrm{Tot}}  : \XX^{\mathrm{Tot}}_{0} \to \XX^{\mathrm{Tot}}_{-1} \big)
~,\qquad \holim(\XX)_n = \XX^{\mathrm{Tot}}_n~,
\end{flalign}
\end{subequations}
for all $n\geq 1$, and the differential is given by $\delta = \delta^{\mathrm{Tot}}$.

\subsection{\label{app:hocolim}Homotopy colimits for non-positively graded chain complexes}
Let $\DD$ be a small category. Given any functor $\YY : \DD \to \Ch_{\leq 0}(\Ab)$,
the homotopy colimit $\hocolim(\YY)$ is constructed in a three step procedure:
First, one takes the simplicial replacement of the diagram $\YY : \DD \to \Ch_{\leq 0}(\Ab)$,
which gives a simplicial object in $\Ch_{\leq 0}(\Ab)$. Then one assigns 
a double chain complex in $\Ch_{\geq 0}(\Ch_{\leq 0}(\Ab))$ via the normalized Moore complex, where the first grading is non-negative and
the second grading is non-positive. Finally one forms the $\coprod$-total complex, 
which gives the homotopy colimit $\hocolim(\YY)$ after truncation to non-positive degrees.
Notice that this is precisely the dual of the construction for homotopy limits presented in Subsection \ref{app:holim}. 
However, we find it useful to go through  these steps in more detail and give explicit formulas.
\sk

Denoting as before the nerve of the small category $\DD$ by $\{\DD_n\}_{n\in\bbN_0}$,
the simplicial replacement of $\YY : \DD \to \Ch_{\leq 0}(\Ab)$ is the simplicial object
in $\Ch_{\leq 0}(\Ab)$ given by
\begin{flalign}\label{eqn:srepY}
\xymatrix{
\coprod\limits_{d\in \DD_0}\, \YY(d)  &  \ar@<0.5ex>[l]\ar@<-0.5ex>[l] 
\coprod\limits_{f\in \DD_1}\, \YY(\mathrm{s}(f)) &  \ar@<1ex>[l]\ar@<-1ex>[l]\ar[l] 
\coprod\limits_{(f_1,f_2)\in \DD_2}\, \YY(\mathrm{s}(f_2)) &
\ar@<0.5ex>[l]\ar@<-0.5ex>[l] \ar@<1.5ex>[l]\ar@<-1.5ex>[l] \cdots~,
}
\end{flalign}
where the arrows are the face maps and we have suppressed 
the degeneracy maps for notational convenience. Here $\coprod$ denotes the coproduct  in the category
$\Ch_{\leq 0}(\Ab)$ and we have denoted by $\mathrm{s}(f)$ the source of a morphism
$f$ in $\DD$.
The face maps $\partial_{i}^n : \coprod_{(f_1,\dots , f_n)\in \DD_n}\, \YY(\mathrm{s}(f_n))\to
\coprod_{(f_1,\dots,f_{n-1})\in \DD_{n-1}}\, \YY(\mathrm{s}(f_{n-1}))$, for $n\geq 1$ and $i=0,1,\dots,n$,
are defined using the universal property of the coproduct; explicitly,
for $i=0$,
\begin{subequations}\label{eqn:faceY}
\begin{flalign}
\xymatrix{
 \coprod\limits_{(f_1,\dots , f_n)\in \DD_n}\, \YY(\mathrm{s}(f_n)) \ar@{.>}[rr]^-{\partial_0^n}&& 
 \coprod\limits_{(f_1,\dots , f_{n-1})\in \DD_{n-1}}\, \YY(\mathrm{s}(f_{n-1})) \\
 &\ar[ul]^-{\iota_{(h_1,\dots,h_n)~~~~}}\YY(\mathrm{s}(h_n)) \ar[ur]_-{~~~~~~~~~~\iota_{(h_2,\dots,h_n)}}&
 }
\end{flalign}
for $1\leq i\leq n-1$, 
\begin{flalign}
\xymatrix{
 \coprod\limits_{(f_1,\dots , f_n)\in \DD_n}\, \YY(\mathrm{s}(f_n)) \ar@{.>}[rr]^-{\partial_i^n}&& 
 \coprod\limits_{(f_1,\dots , f_{n-1})\in \DD_{n-1}}\, \YY(\mathrm{s}(f_{n-1})) \\
 &\ar[ul]^-{\iota_{(h_1,\dots,h_n)~~~~}}\YY(\mathrm{s}(h_n)) \ar[ur]_-{~~~~~~~~~~\iota_{(h_1,\dots,h_i\circ h_{i+1},\dots,h_n)}}&
 }
\end{flalign}
and for $i=n$,
\begin{flalign}
\xymatrix{
 \coprod\limits_{(f_1,\dots , f_n)\in \DD_n}\, \YY(\mathrm{s}(f_n)) \ar@{.>}[rr]^-{\partial_n^n}&& 
 \coprod\limits_{(f_1,\dots , f_{n-1})\in \DD_{n-1}}\, \YY (\mathrm{s}(f_{n-1})) \\
 \ar[u]^-{\iota_{(h_1,\dots,h_n)}~}\YY(\mathrm{s}(h_n)) \ar[rr]_-{\YY(h_n)}&& \YY(\mathrm{s}(h_{n-1}))\ar[u]_-{~\iota_{(h_1,\dots,h_{n-1})}}
 }
\end{flalign}
\end{subequations}
where $\iota_{-}$ are the inclusion morphisms to the coproducts.  For $n\geq 0$ and $i=0,1,\dots,n$, the degeneracy maps $\epsilon_i^n : \coprod_{(f_1,\dots , f_n)\in \DD_n}\, \YY(\mathrm{s}(f_n))\to
\coprod_{(f_1,\dots,f_{n+1})\in \DD_{n+1} }\,\YY(\mathrm{s}(f_{n+1}))$ 
are also defined by using the universal property of the coproduct; explicitly,
for all $i=0,1,\dots,n$,
\begin{flalign}\label{eqn:degY}
\xymatrix{
 \coprod\limits_{(f_1,\dots , f_n)\in \DD_n}\, \YY(\mathrm{s}(f_n)) \ar@{.>}[rr]^-{\epsilon_i^n}&& 
 \coprod\limits_{(f_1,\dots , f_{n+1})\in \DD_{n+1}}\, \YY(\mathrm{s}(f_{n+1})) \\
 &\ar[ul]^-{\iota_{(h_1,\dots,h_n)~~~~}}\YY(\mathrm{s}(h_n)) \ar[ur]_-{~~~~~~~~~~~\iota_{(h_1,\dots,h_i,\id,h_{i+1},\dots,h_n)}}&
 }
\end{flalign}

Using the normalized Moore complex (\ref{eqn:Moorecomplex}), we can assign to the simplicial object (\ref{eqn:srepY})
in $\Ch_{\leq 0}(\Ab)$ a double chain complex in $\Ch_{\geq 0}(\Ch_{\leq 0}(\Ab))$.
Denoting this double chain complex by $\YY_{\ast,\ast}$, a simple calculation shows that
\begin{flalign}\label{eqn:chainchainY}
\YY_{0,\ast} = \coprod\limits_{d\in \DD_0}\, \YY(d)~,\qquad
\YY_{n,\ast} = 
\coprod\limits_{\mycom{(f_1,\dots,f_n)\in \DD_n}{f_i\neq \id }}\, \YY(\mathrm{s}(f_n))~,
\end{flalign}
for all $n\geq 1$, where the second coproduct is taken over all composable $n$-arrows $(f_1,\dots,f_n)$
such that none of the $f_i$ is an identity morphism.
The vertical differential $\delta^\mathrm{v}  : \YY_{\ast,\ast} \to \YY_{\ast-1,\ast}$
is given by the alternating sum of the face maps, i.e.\ $\delta^\mathrm{v} = \sum_{i=0}^{n}\, (-1)^i~\partial^n_i$
on $\YY_{n,\ast}$, and the horizontal differential $\delta^\mathrm{h}  : 
\YY_{\ast,\ast} \to\YY_{\ast,\ast-1}$ is given by the coproduct of
the differentials in the chain complexes $\YY(d)$, for $d$ an object in $\DD$. The double complex
$\YY_{\ast,\ast}$ may be visualized as
\begin{flalign}\label{eqn:doubleY}
\xymatrix{
\ddots&&\vdots\ar[d]_-{\delta^{\mathrm{v}}} && \ar[d]_-{\delta^{\mathrm{v}}}\vdots&&\vdots \ar[d]^-{\delta^{\mathrm{v}}}\\
\cdots&&\ar[ll]_-{\delta^{\mathrm{h}}} \YY_{2,-2} \ar[d]_-{\delta^{\mathrm{v}}}&& \ar[ll]_-{\delta^{\mathrm{h}}}\ar[d]_-{\delta^{\mathrm{v}}}\YY_{2,-1} && \YY_{2,0}\ar[ll]_-{\delta^{\mathrm{h}}}\ar[d]^-{\delta^{\mathrm{v}}}\\
\cdots&&\ar[ll]_-{\delta^{\mathrm{h}}} \YY_{1,-2}\ar[d]_-{\delta^{\mathrm{v}}}&& \ar[ll]_-{\delta^{\mathrm{h}}}\ar[d]_-{\delta^{\mathrm{v}}}\YY_{1,-1} && \YY_{1,0}\ar[ll]_-{\delta^{\mathrm{h}}}\ar[d]^-{\delta^{\mathrm{v}}}\\
\cdots&&\ar[ll]_-{\delta^{\mathrm{h}}} \YY_{0,-2}&&\ar[ll]_-{\delta^{\mathrm{h}}} \YY_{0,-1} && \YY_{0,0}\ar[ll]_-{\delta^{\mathrm{h}}}
}
\end{flalign}

We now form the $\coprod$-total complex
\begin{flalign}\label{eqn:totalY}
\YY^{\mathrm{Tot}} :=\Big(\bigoplus_{n\in\bbZ}\, \YY^{\mathrm{Tot}}_n\,,\delta^{\mathrm{Tot}}\Big) 
:=  \Big(\bigoplus_{n\in\bbZ} \ \coprod_{p+q=n}\, \YY_{p,q}\,,~\delta^{\mathrm{Tot}} := \delta^{\mathrm{v}} + (-1)^p\,\delta^{\mathrm{h}} \Big)
\end{flalign}
and we notice that $\YY^{\mathrm{Tot}}$ is a {\em $\bbZ$-graded} chain complex of Abelian groups, in particular
it is non-trivial in positive degrees.
The homotopy colimit $\hocolim(\YY)$ of the diagram $\YY :\DD\to\Ch_{\leq 0}(\Ab)$ is then the truncation
of $\YY^{\mathrm{Tot}}$ to non-positive degrees. Explicitly,
\begin{subequations}\label{eqn:hocolimY}
\begin{flalign}
\hocolim(\YY) = \Big(\bigoplus_{n\leq 0} \, \hocolim(\YY)_n\,,~\delta\Big) ~,
\end{flalign}
where
\begin{flalign}
\hocolim(\YY)_0 = \frac{\YY^{\mathrm{Tot}}_0}{\Imm\big(\delta^{\mathrm{Tot}}  : \YY^{\mathrm{Tot}}_{1} \to 
\YY^{\mathrm{Tot}}_{0} \big)}~,
\qquad \hocolim(\YY)_{n} = \YY^{\mathrm{Tot}}_{n}~,
\end{flalign}
\end{subequations}
for all $n\leq - 1$, and the differential is given by $\delta = \delta^{\mathrm{Tot}}$.



\begin{thebibliography}{10}

\bibitem[ACMM86]{Abbati:1986ww} 
  M.~C.~Abbati, R.~Cirelli, A.~Mania and P.~Michor,
  ``Smoothness of the action of the gauge transformation group on connections,''
  J.\ Math.\ Phys.\  {\bf 27} (1986) 2469.
  
\bibitem[ACM89]{ACM}
M.~C.~Abbati, R.~Cirelli and A.~Mania,
``The orbit space of the action of the gauge transformation group on connections,''
J.\ Geom.\ Phys.\ {\bf 6} (1989) 537.   
  
\bibitem[BSS14]{Becker:2014tla} 
  C.~Becker, A.~Schenkel and R.~J.~Szabo,
  ``Differential cohomology and locally covariant quantum field theory,''
  arXiv:1406.1514 [hep-th].

\bibitem[BM06]{Belov:2006jd}
  D.~Belov and G.~W.~Moore,
  ``Holographic action for the self-dual field,''
  arXiv:hep-th/0605038.

\bibitem[Ben14]{Benini:2014vsa}
  M.~Benini,
  ``Optimal space of linear classical observables for Maxwell $k$-forms via spacelike and timelike compact de Rham cohomologies,''
  arXiv:1401.7563 [math-ph].
 
\bibitem[BDHS14]{Benini:2013ita}
  M.~Benini, C.~Dappiaggi, T.-P.~Hack and A.~Schenkel,
  ``A $C^*$-algebra for quantized principal $U(1)$-connections on globally hyperbolic Lorentzian manifolds,''
  Commun.\ Math.\ Phys.\  {\bf 332} (2014) 477
  [arXiv:1307.3052 [math-ph]].

\bibitem[BDS14]{Benini:2013tra}
  M.~Benini, C.~Dappiaggi and A.~Schenkel,
  ``Quantized Abelian principal connections on Lorentzian manifolds,''
  Commun.\ Math.\ Phys.\  {\bf 330} (2014) 123
  [arXiv:1303.2515 [math-ph]]. 

\bibitem[Bou10]{Bouwknegt:2010zz} 
  P.~Bouwknegt,
  ``Lectures on cohomology, T-duality, and generalized geometry,''
  Lect.\ Notes Phys.\  {\bf 807} (2010) 261.
  
\bibitem[BFR13]{Brunetti:2013maa}
  R.~Brunetti, K.~Fredenhagen and K.~Rejzner,
  ``Quantum gravity from the point of view of locally covariant quantum field theory,''
  arXiv:1306.1058 [math-ph].
  
\bibitem[BFR12]{Brunetti:2012ar} 
  R.~Brunetti, K.~Fredenhagen and P.~L.~Ribeiro,
  ``Algebraic structure of classical field theory I: Kinematics and linearized dynamics for real scalar fields,''
  arXiv:1209.2148 [math-ph].

\bibitem[BFV03]{Brunetti:2001dx}
  R.~Brunetti, K.~Fredenhagen and R.~Verch,
  ``The generally covariant locality principle: A new paradigm for local quantum field theory,''
  Commun.\ Math.\ Phys.\  {\bf 237} (2003) 31
  [arXiv:math-ph/0112041].
  
\bibitem[Bry07]{Brylinski} 
  J.-L.\ Brylinski, 
  ``Loop Spaces, Characteristic Classes and Geometric  Quantization,''
  Birkh\"auser, Boston (2007).
  
\bibitem[CC04]{CC}
J.~L.~Castiglioni and G.~Corti$\tilde{\text{n}}$as,
``Cosimplicial versus DG-rings: A version of the Dold-Kan correspondence,''
J.\ Pure Appl. Algebra {\bf 191} (2004) 119
[arXiv:math.KT/0306289].

\bibitem[CRV12]{Ciolli:2011xv} 
  F.~Ciolli, G.~Ruzzi and E.~Vasselli,
  ``Causal posets, loops and the construction of nets of local algebras for QFT,''
  Adv.\ Theor.\ Math.\ Phys.\  {\bf 16} (2012) 645 
  [arXiv:1109.4824 [math-ph]].
  
\bibitem[CRV13]{Ciolli:2013pta} 
  F.~Ciolli, G.~Ruzzi and E.~Vasselli,
  ``QED representation for the net of causal loops,''
  arXiv:1305.7059 [math-ph].
  
\bibitem[Cra03]{Crainic}
    M.~Crainic, 
    ``Differentiable and algebroid cohomology, van {E}st
      isomorphisms, and characteristic classes,''
       Comment.\ Math.\ Helv.\ {\bf 78} (2003) 681
       [arXiv:math.DG/0008064].
  
\bibitem[DL12]{Dappiaggi:2011zs} 
  C.~Dappiaggi and B.~Lang,
  ``Quantization of Maxwell's equations on curved backgrounds and general local covariance,''
  Lett.\ Math.\ Phys.\  {\bf 101} (2012) 265
  [arXiv:1104.1374 [gr-qc]].
  
\bibitem[DS13]{Dappiaggi:2011cj}
  C.~Dappiaggi and D.~Siemssen,
  ``Hadamard states for the vector potential on asymptotically flat spacetimes,''
  Rev.\ Math.\ Phys.\  {\bf 25} (2013) 1350002
  [arXiv:1106.5575 [gr-qc]].
  
\bibitem[Dug]{Dugger}
D.~Dugger, 
``A primer on homotopy colimits,''
available at \url{http://pages.uoregon.edu/ddugger/hocolim.pdf}.      
      
\bibitem[DS95]{Dwyer}
     W.~G.~Dwyer and J.~Spalinski,
``Homotopy theories and model categories,''
 in: {\sl Handbook of Algebraic Topology}, North-Holland, Amsterdam
 (1995) 73. 

\bibitem[Fan01]{Fantechi}
B.~Fantechi, 
``Stacks for everybody,''
Progr.\ Math. {\bf 201} (2001) 349.

\bibitem[FL14]{Fewster:2014hba} 
  C.~J.~Fewster and B.~Lang,
  ``Dynamical locality of the free Maxwell field,''
  arXiv:1403.7083 [math-ph].

\bibitem[FP03]{Fewster:2003ey}
  C.~J.~Fewster and M.~J.~Pfenning,
  ``A quantum weak energy inequality for spin one fields in curved space-time,''
  J.\ Math.\ Phys.\  {\bf 44} (2003) 4480
  [arXiv:gr-qc/0303106].
  
\bibitem[FSS15]{Fiorenza:2013jz}
  D.~Fiorenza, H.~Sati and U.~Schreiber,
  ``A higher stacky perspective on Chern-Simons theory,''
  in: {\sl Mathematical Aspects of Quantum Field Theories}, D.~Calaque and T.~Strobl, eds., Mathematical Physics Studies, Springer (2015) 153
  [arXiv:1301.2580 [hep-th]].
  
\bibitem[Fre90]{Fre90}
K.~Fredenhagen, 
``Generalizations of the theory of superselection sectors,''
in: {\sl The Algebraic Theory of Superselection Sectors: Introduction and
  Recent Results}, D.~Kastler, ed., World
  Scientific Publishing (1990) 379.

\bibitem[Fre93]{Fre93}
K.~Fredenhagen, 
``Global observables in local quantum physics,''
 in: {\sl Quantum and Non-Commutative Analysis: Past, Present and Future Perspectives},
 H.~Araki, K.~R.~Ito, A.~Kishimoto, and I.~Ojima,
  eds., Kluwer Academic Publishers (1993) 41.

\bibitem[FRS92]{FRS92}
K.~Fredenhagen, K.-H.~Rehren and B.~Schroer, 
``Superselection sectors with braid group statistics and exchange algebras II: Geometric aspects and
  conformal covariance,'' 
  Rev.\ Math.\ Phys.\ {\bf 4}
  (1992) 113.

\bibitem[FR12]{Fredenhagen:2011an}
  K.~Fredenhagen and K.~Rejzner,
  ``Batalin-Vilkovisky formalism in the functional approach to classical field theory,''
  Commun.\ Math.\ Phys.\  {\bf 314} (2012) 93
  [arXiv:1101.5112 [math-ph]].

\bibitem[FR13]{Fredenhagen:2011mq}
  K.~Fredenhagen and K.~Rejzner,
  ``Batalin-Vilkovisky formalism in perturbative algebraic quantum field theory,''
  Commun.\ Math.\ Phys.\  {\bf 317} (2013) 697
  [arXiv:1110.5232 [math-ph]].

\bibitem[Fre00]{Freed:2000ta}
  D.~S.~Freed,
  ``Dirac charge quantization and generalized differential cohomology,''
  Surv. Diff. Geom. {\bf VII} (2000) 129
  [arXiv:hep-th/0011220].

\bibitem[FW99]{Freed:1999vc}
  D.~S.~Freed and E.~Witten,
  ``Anomalies in string theory with D-branes,''
  Asian J.\ Math {\bf 3} (1999) 819
  [arXiv:hep-th/9907189].

\bibitem[GJ99]{GJ99}
P.~G.~Goerss and J.~F.~Jardine,
``Simplicial Homotopy Theory,''
Birkh\"auser Verlag, Basel (1999).

\bibitem[Hol08a]{Hollander}
S.~Hollander,
``A homotopy theory for stacks,''
Israel J. Math.\ {\bf 163} (2008) 93
[arXiv:math.AT/0110247].

\bibitem[Hol08b]{Hollands:2007zg}
  S.~Hollands,
  ``Renormalized quantum Yang-Mills fields in curved spacetime,''
  Rev.\ Math.\ Phys.\  {\bf 20} (2008) 1033
  [arXiv:0705.3340 [gr-qc]].

\bibitem[HS05]{HS}
M.~J.~Hopkins and I.~M.~Singer,
  ``Quadratic functions in geometry, topology, and M-theory,''
  J.\ Diff.\ Geom.\  {\bf 70} (2005) 329
  [arXiv:math.AT/0211216].

\bibitem[Jar97]{Jardine}
J.~F.~Jardine, 
``A closed model structure for differential graded algebras,'' 
Fields Inst. Commun. {\bf 17} (1997) 55.

\bibitem[Kha14]{Khavkine:2014kya}
  I.~Khavkine,
  ``Covariant phase space, constraints, gauge and the Peierls formula,''
  Int.\ J.\ Mod.\ Phys.\ A {\bf 29} (2014) 1430009
  [arXiv:1402.1282 [math-ph]].

\bibitem[Kha15]{Khavkine:2015fwa}
  I.~Khavkine,
  ``Local and gauge invariant observables in gravity,''
  arXiv:1503.03754 [gr-qc].

\bibitem[Rod14]{RodGon}
B.~Rodr{\'i}guez-Gonz{\'a}lez,
``Realizable homotopy colimits,''
Theor. Appl. Cat. {\bf 29} (2014) 609
[arXiv:1104.0646 [math.AG]].

\bibitem[SDH14]{Sanders:2012sf}
  K.~Sanders, C.~Dappiaggi and T.-P.~Hack,
  ``Electromagnetism, local covariance, the Aharonov-Bohm effect and Gauss' law,''
  Commun.\ Math.\ Phys.\  {\bf 328} (2014) 625
  [arXiv:1211.6420 [math-ph]].

\bibitem[Sza12]{Szabo:2012hc} 
   R.~J.~Szabo,
   ``Quantization of higher Abelian gauge theory in generalized differential cohomology,''
   PoS ICMP {\bf 2012} (2012) 009
   [arXiv:1209.2530 [hep-th]].
   
\bibitem[Vis05]{Vistoli}   
A.~Vistoli,
``Grothendieck topologies, fibred categories and descent theory,''
Math.\ Surv. Monogr. {\bf 123} (2005) 1
[arXiv:math.AG/0412512].

\end{thebibliography}
\end{document}